\documentclass[12pt]{article}

\usepackage{newtxtext,newtxmath}
\usepackage{graphicx}

\usepackage[letterpaper,margin=1in]{geometry}

\renewenvironment{abstract}
	{\quotation}
	{\endquotation}

\date{}

\makeatletter
\renewcommand{\fnum@figure}{\textbf{Figure \thefigure}}
\renewcommand{\fnum@table}{\textbf{Table \thetable}}
\makeatother

\usepackage{scicite}

\usepackage{url}

\def\scititle{
	Charge-transfer excitation drives cooperative phonon-assisted exciton self-trapping in NiPS$_3$
}
\title{\bfseries \boldmath \scititle}

\author{
	Long~Cheng$^{1\ast}$,
	Alexander~Holm~Ohrt$^{1}$,
	Oliver~Nagy$^{1}$,
	Yu~Zhao$^{2,3}$\and
	Chenyu~Wang$^{4}$,
	Yaxian~Wang$^{4}$,
	Sheng~Meng$^{4}$,
	Guohui~Li$^{5}$,
	Qinying~Pan$^{6,7}$\and
	Xingyu~Huang$^{1}$,
	Sanshui~Xiao$^{1}$,
	Sijie~Zhang$^{4}$,
	Guoqian~Liao$^{4}$,
	Yutong~Li$^{4}$\and
	Xuan~Luo$^{2}$,
	Yuping~Sun$^{2,8,9}$,
	Kaibo~Zheng$^{7}$,
	Peter~Uhd~Jepsen$^{1}$,
	Binbin~Zhou$^{1\ast}$\and
	\small$^{1}$Department of Electrical and Photonics Engineering,\and
	\small Technical University of Denmark, Kgs. Lyngby, Denmark.\and
	\small$^{2}$Key Laboratory of Materials Physics, Institute of Solid State Physics,\and
	\small HFIPS, Chinese Academy of Sciences, Hefei, China.\and
	\small$^{3}$Science Island Branch of Graduate School, University of Science and Technology of China, Hefei, China.\and
	\small$^{4}$Beijing National Laboratory for Condensed Matter Physics,\and
	\small Institute of Physics, Chinese Academy of Sciences, Beijing, China.\and
	\small$^{5}$College of Physics and Optoelectronics, Taiyuan University of Technology, Taiyuan, China.\and
	\small$^{6}$Department of Chemistry, Technical University of Denmark, Kgs. Lyngby, Denmark.\and
    \small$^{7}$Department of Chemical Physics and NanoLund Chemical Center, Lund University, Lund, Sweden.\and
	\small$^{8}$Anhui Provincial Key Laboratory of Low-Energy Quantum Materials and Devices,\and
	\small High Magnetic Field Laboratory, HFIPS, Chinese Academy of Sciences, Hefei, China.\and
	\small$^{9}$Collaborative Innovation Center of Advanced Microstructures, Nanjing University, Nanjing, China.\and
	\small$^\ast$Corresponding authors. Email: longch@dtu.dk; zhou@dtu.dk
}

\begin{document}

\maketitle

% Abstract, in bold
\begin{abstract} \bfseries \boldmath
Carrier self-trapping is a fundamental relaxation pathway in strongly correlated quantum materials, yet the microscopic lattice motions that drive it have remained largely inaccessible to direct observation. Here we investigate the van der Waals antiferromagnet NiPS$_3$ upon excitation across its ligand-to-metal charge-transfer gap, and reveal a nonequilibrium pathway that initially generates unbound carriers before they evolve into correlated excitonic states. Using ultrabroadband optical-pump terahertz-probe spectroscopy, we simultaneously track carrier localization and the associated lattice dynamics in real time. The transient conductive response quenches with a time constant of $\sim$109 fs and evolves into a localized exciton state, accompanied by pronounced renormalization of Ni--S optical phonons and the emergence of a transient 2.7~THz vibrational mode. Temperature-dependent measurements, together with  mode-resolved dielectric dynamics and symmetry analysis, identify the 2.7 and 8.0~THz responses as spectroscopic indicators of exciton self-trapping and point to a cooperative two-phonon mechanism. These findings establish the excitation pathway as a key parameter controlling carrier localization and many-body relaxation in strongly coupled quantum materials.
\end{abstract}

% Keywords from the IOP version. The SA template has no keyword field, so they are not typeset:
% \keywords{Self-trapped excitons; charge-transfer excitation; cooperative phonon coupling; ultrafast terahertz spectroscopy; van der Waals antiferromagnets}

\section*{Introduction}\label{sec1}
In correlated quantum materials, the competition between electronic delocalization and lattice-driven carrier localization plays a central role in determining nonequilibrium dynamics. Optical excitation can drive systems with strongly coupled charge, spin, orbital, and lattice degrees of freedom far from equilibrium, generating transient states inaccessible in equilibrium \cite{multiDegree, lightDriveLattice, non-adiabatic-dynamics}. The relaxation of photoexcited carriers in strongly coupled materials is governed by the interplay between electronic and lattice degrees of freedom, which drives carrier localization, exciton formation, and emergent many-body behavior \cite{polaron, excitonControl, exciton}. Yet the microscopic lattice coordinates that mediate these transformations are rarely identified directly. Determining such reaction coordinates is therefore a prerequisite for understanding and controlling nonequilibrium phenomena in correlated quantum materials.

Two-dimensional van der Waals magnets are particularly well suited to such studies because the reduced dimensionality enhances Coulomb interactions and amplifies the coupling between charge, spin, and lattice degrees of freedom \cite{2Dmagnets1,2Dmagnets2,2Dmagnets3,manybody}. Within this class, the charge-transfer antiferromagnet NiPS$_3$ offers an ideal setting in which strong S $3p$ -- Ni $3d$ hybridization makes the elementary excitations highly sensitive to local lattice distortions\cite{phyNPS}. Optical and Raman spectroscopies have revealed pronounced exciton--phonon coupling through the observation of phonon sidebands, exciton--phonon bound states, and lattice-induced spectral renormalization\cite{excitonphonon,elecphonon}, highlighting the central role of lattice dynamics in shaping the excited-state landscape of NiPS$_3$.

Recent ultrafast studies have reported diverse nonequilibrium responses in NiPS$_3$ following resonant or near-resonant excitation of the excitonic manifold ($\sim$1.47--1.55 eV), ranging from transient Drude-like conductivity \cite{RN3556} to population-inverted excitonic states accompanied by stimulated emission \cite{negsig}. However, these studies primarily probe low-energy excitations that directly address pre-existing excitonic or localized $d$--$d$ states. By contrast, excitation across the ligand-to-metal charge-transfer (LMCT) gap generates initially unbound, high-energy carriers and accesses a qualitatively different nonequilibrium relaxation pathway. At the microscopic level, the LMCT transition transfers electrons from sulfur $3p$ orbitals into nickel $3d$ states, inducing substantial inter-site charge redistribution and strongly perturbing the local Ni--S bonding environment \cite{phyNPS,abinitio}. Unlike resonant excitation of pre-existing excitons, this process injects substantial excess energy into the coupled charge--lattice system and strongly perturbs the local Ni--S bonding environment, driving pronounced non-adiabatic dynamics. The nonequilibrium trajectory of the photoexcited state is then determined by the competition between electronic delocalization and lattice-assisted localization. Time-resolved optical studies indicate that charge-transfer excitations relax into localized excitonic states within $\sim$2~ps \cite{excitonlife}. However, purely electronic probes are inherently insensitive to the lattice coordinates involved. Resolving this mechanism requires simultaneous access to both electronic and lattice dynamics.

Here, we employ sub-100-fs ultrabroadband optical-pump terahertz-probe (OPTP) spectroscopy to map the coupled charge and lattice dynamics of NiPS$_3$ following excitation across the LMCT gap. The exceptionally broad probe bandwidth, extending to 15 THz, provides simultaneous access to macroscopic carrier transport and the optical phonons involved in energy dissipation and lattice relaxation. We observe a rapid quenching of the transient conductive response into a localized exciton state, accompanied by pronounced renormalization of Ni--S optical phonons and the transient activation of a folded low-frequency ($\sim$2.7~THz) vibrational mode superimposed on an imaginary dielectric sign-reversing electronic background. Combined with transient absorption spectroscopy and temperature-dependent OPTP measurements, these excited-state dynamic signatures point to a cooperative phonon-assisted pathway toward self-trapped exciton formation.

\section*{Results}\label{sec2}

\textbf{Crystal structure and static properties of NiPS$_3$.} As illustrated in Fig. \ref{fig:fig1.png}\textbf{A}, NiPS$_3$ crystallizes in a monoclinic ($C2/m$) structure, in which Ni$^{2+}$ ($3d^8$) ions form a honeycomb network coordinated by edge-sharing sulfur octahedra. The local ligand field is trigonally distorted from $O_h$ to $D_{3d}$ symmetry, strongly modifying the crystal-field manifold of the Ni $3d$ orbitals and thereby shaping the low-energy electronic excitations of the material\cite{NPS_Struc, ddtransition3, RN3589}. The diffuse-reflectance absorbance spectrum (Fig. \ref{fig:fig1.png}\textbf{B}) reveals several characteristic optical transitions. Weak features below $1.2$ eV arise from Laporte-forbidden on-site $d$--$d$ excitations from the ${^3}A_{2g}$ ground state into the split ${^3}A_{1g}$ and ${^3}E_g$ states derived from the ${^3}T_{2g}$ manifold\cite{Kim:RN3583}. At higher energies, a broad absorption continuum from $\sim1.2$ to 2.1 eV dominates the optical response. Within this continuum, a pronounced resonance centered near 1.4 eV (A) is consistent with the spin-orbit-entangled charge-transfer excitons reported previously in NiPS$_3$ \cite{RN3556,ddtransition1}. Although substantially broadened at room temperature owing to enhanced thermal decoherence and the absence of long-range magnetic stabilization\cite{Kim:RN3583}, the persistence of this feature indicates that excitonic correlations remain significant in the paramagnetic phase.

The strongest absorption feature, centered at $\sim$1.6 eV (B), originates from a ligand-to-metal charge-transfer (LMCT) transition\cite{CTgap300K}. Density-functional-theory (DFT) calculations (fig. \ref{fig:figS2.png}) show that this excitation transfers electrons from S $3p$ into antibonding Ni $3d$ ($e_g$) states, producing inter-site redistribution of electronic charge within the Ni--S network. Moreover, within the strong-coupling framework\cite{CTmaterial_e-phon_couple}, the narrow bandwidth of the localized Ni $3d$ states enhances electron-phonon coupling, making LMCT excitation a particularly effective route to carrier localization irrespective of long-range magnetic order \cite{RN3592}. Additional absorption bands near 1.78 and 2.0 eV are assigned to higher-energy crystal-field ($d$--$d$) excitations associated with the split ${^3}T_{1g}(F)$ manifold (Supplementary fig. \ref{fig:figS3.png}). We therefore employ a 1.91 eV (650 nm) pump, which lies above the optical gap to access the LMCT continuum rather than enter the low-energy excitonic manifold or the Laporte-forbidden crystal-field transitions, whose oscillator strengths are orders of magnitude weaker. This excitation pathway generates initially unbound carriers and injects substantial excess energy into the coupled charge-lattice system, thereby accessing a nonequilibrium regime distinct from that explored in previous resonant-excitation studies.

The lattice degrees of freedom are characterized by the THz transmission spectrum shown in Fig. \ref{fig:fig1.png}\textbf{C}. Two prominent infrared-active optical phonons are observed at 5.7 and 8.0 THz. In particular, the high energy mode corresponds to an odd-parity A$_u$ vibration involving in-plane sulfur motions that modulate the Ni--S octahedral environment and the crystal-field splitting of the Ni $d$-levels, whereas the 5.7 THz mode is a B$_u$ vibration dominated by Ni displacements within the NiS$_6$ octahedra (Supplementary fig. \ref{fig:figS4.png}\textbf{A})\cite{ir_active,ir_active1,phonons}. Exciton–phonon coupling is independently evidenced in the room-temperature photoluminescence (PL) spectrum (Fig. \ref{fig:fig1.png}\textbf{D}), which shows two partially resolved excitonic emission features, A$_1$ and A$_2$, separated by $E_s\approx46.0 \pm 0.5$ meV. Their relative spectral weights shift systematically with excitation power, consistent with previous reports of pronounced exciton-phonon coupling in NiPS$_3$\cite{excitonphonon}. At higher excitation densities, both excitonic branches exhibit sub-linear saturation together with the appearance of a lower-energy sideband (A$_3$) near 1.295 eV (Supplementary fig. \ref{fig:figS5.png}), pointing to additional many-body relaxation channels such as exciton-exciton annihilation. Together, these measurements establish NiPS$_3$ as a strongly coupled charge-lattice system and define the spectroscopic reference frame for the ultrafast experiments.

\noindent\textbf{Characterization of ultrafast excitonic dynamics.}
To investigate the nonequilibrium evolution of charge-transfer excitations in NiPS$_3$, we employ ultrabroadband optical-pump terahertz-probe (OPTP) spectroscopy in the reflection geometry shown in Fig.~\ref{fig:fig2.png}\textbf{A}. This configuration avoids strong attenuation of the probe field by the intrinsic infrared-active phonons (Fig.~\ref{fig:fig1.png}\textbf{C}), thereby preserving the ultrabroad bandwidth and enabling direct monitoring of the transient electrodynamic response. Following 650 nm (1.91 eV) excitation at a fluence of 1.3 mJ/cm$^2$, the transient reflectivity change ($\Delta R$) rises rapidly with a characteristic time ($\tau$) of $\sim$90 fs (Fig.~\ref{fig:fig2.png}\textbf{B}), marking the prompt generation of itinerant carriers in the LMCT continuum. This initial response then decays with $\tau \approx 109$ fs, far faster than typical carrier recombination or lattice thermalization processes\cite{RN3556,MoS2_exciton_timescale,WSe2_exciton_timescale}. Such rapid suppression of the conductive response indicates efficient conversion of initially photoexcited electron-hole pairs into strongly coupled excitons. Following this stage, the THz response undergoes a sign inversion and evolves into a pronounced negative signal over $\sim$3 ps. This behavior contrasts sharply with the monotonic recovery commonly observed in conventional semiconductors and indicates a qualitative change in the nature of the photoexcited state. The negative response signifies that the transient electrodynamics are no longer governed by itinerant carriers but instead by localized photoexcited states with a predominantly dielectric character. On longer timescales, the signal gradually recovers with a characteristic lifetime of $\tau \approx 29$ ps.

Complementary transient absorption spectroscopy (TAS) measurements (Fig.~\ref{fig:fig2.png}\textbf{B}, black symbols) provide an independent probe of the excited-state population dynamics underlying the transient electrodynamic response. Following an initial $\sim$1~ps build-up of the band-edge ground-state bleach, the TAS signal exhibits a fast decay with a characteristic time of $\sim$2.7 ps, closely matching the timescale over which the negative THz response develops. The bleach build-up, slower than the prompt THz rise and steepening as it proceeds, suggests transient occupation of a charge-neutral intermediate state such as an on-site $d$--$d$ excitation. Analysis of the TAS amplitudes indicates that approximately 60$\%$ of the photoexcited population is transferred into a long-lived localized excitonic state. Notably, this fraction is comparable to the relative spectral weight ($A_2/A_1$) observed in the steady-state PL spectrum (Fig.~\ref{fig:fig1.png}\textbf{D}). This development, synchronized with the negative THz trajectory, indicates their direct connection and is consistent with the conversion of a large fraction of the initially photoexcited carriers into localized excitonic configurations. The remaining excitonic population relaxes on a nanosecond timescale ($\sim$1.6~ns).

The origin of the THz response is further elucidated by pump-fluence-dependent measurements. As shown in Fig.~\ref{fig:fig2.png}\textbf{C,D}, both the initial positive peak and the subsequent negative minimum scale linearly with excitation fluence across the investigated range. The absence of saturation or threshold behavior indicates that the dynamics are not dominated by high-density many-body effects. In particular, the linearity rules out strong exciton--exciton annihilation and is inconsistent with the threshold-like behavior expected for stimulated emission or population inversion scenarios proposed under resonant excitation conditions\cite{negsig,RN3556}. Instead, the results point to an intrinsic localization process that remains operative over a broad range of excitation densities. Temperature-dependent results provide further insight, as shown in Fig.~\ref{fig:fig2.png}\textbf{E}. Lowering the temperature progressively suppresses the negative response and slows the localization kinetics, a signature of thermal activation. Arrhenius analysis of the negative amplitude (Fig.~\ref{fig:fig2.png}\textbf{E}, inset) and the corresponding build-up rate (fig.~\ref{fig:figS6.png}\textbf{B}) yields two numerically close yet physically distinct energies, $E_A\approx43\pm2.3$ meV and $E_b\approx45\pm2$ meV, respectively. $E_A$, consistent with $E_s$ from the PL spectrum, characterizes the thermally activated growth of the localized population, whereas $E_b$ is the activation barrier between the free-exciton and localized states. $E_b$ of this magnitude places NiPS$_3$ in the shallow self-trapping regime, in which free and localized excitons remain simultaneously accessible at room temperature rather than the photoexcited population collapsing irreversibly into a single deep minimum (Supplementary \ref{si:1dt}). It thus sets a quantitative benchmark for the microscopic origin of the localized state.

\noindent\textbf{Spectro-temporal mapping and phonon-resolved dynamics.} 
To directly probe the lattice dynamics accompanying exciton localization, we extracted the transient complex dielectric function ($\Delta\tilde{\varepsilon}$) from the two-dimensional spectro-temporal OPTP measurements (Fig. \ref{fig:fig3.png}\textbf{A, B}, details in Supplementary \ref{si:pgfm_and_ptmm}). The real component ($\Delta\varepsilon_{\mathrm{re}}$) reflects changes in dipole polarizability, whereas the imaginary component ($\Delta\varepsilon_{\mathrm{im}}$) tracks energy dissipation in the coupled charge--lattice system. Immediately after photoexcited maximum ($t=0$), $\Delta\varepsilon_{\mathrm{im}}$ shows a short-lived positive response during the first $\sim$109~fs window, corresponding to the itinerant-carrier-dominated nonequilibrium regime. Subsequently, the macroscopic low-frequency ($<$5 THz) $\Delta\varepsilon_{\mathrm{im}}$ rapidly drops to negative and reaches the minimum at $\sim$3~ps, while the real dielectric response $\Delta\varepsilon_{\mathrm{re}}$ undergoes a concurrent rise. This delayed buildup, coinciding with the characteristic localization timescale identified from the OPTP kinetics signals the prompt quenching of the itinerant carrier response and formation of a localized dielectric state.

The intrinsic infrared-active phonons at 5.7 THz ($B_u$) and 8.0 THz ($A_u$) exhibit distinct nonequilibrium dynamics (fig. \ref{fig:figS8.png} and Fig. \ref{fig:fig3.png}\textbf{C}). Unlike the low-frequency modes that undergo an electrodynamic inversion, the $\Delta\varepsilon_{\mathrm{im}}$ responses of both intrinsic modes remain positive. However, the 5.7 THz mode sustains a relative steady amplitude, whereas the 8.0 THz mode undergoes a 70$\%$ suppression during the precursor-exciton formation. Following this window, the dissipative imaginary response ($\Delta\varepsilon_{\mathrm{im}}$) of both modes undergoes a prompt frequency blueshift (Fig. \ref{fig:fig3.png}\textbf{B}). This dynamic stiffening indicates the subsequent spatial contraction of the excitonic wavepacket, exerting localized electrostrictive stress on the NiS$_6$ octahedra and transiently enhancing the effective Ni--S restoring forces. Over the subsequent $\sim$3 ps trajectory, the blueshift persists without progressive lattice softening, signaling a stabilized structural reorganization that accommodates the localized excitonic state\cite{polaron,ExcitonInPhonon}. Concurrently, two transiently brightened modes emerge at $\sim$2.7 and 4.5 THz. Localization of the excitation removes the C-centering translation of the lattice at its site, which amounts to a local doubling of the primitive cell. This folds the zone-boundary dark branches, both transforming as odd-parity modes, to the $\Gamma$ ($q\approx0$) point and relaxes selection constraints, rendering them optically active \cite{ExcitonInPhonon}. Among all these modes (Fig. \ref{fig:fig3.png}\textbf{D}), the 2.7 THz branch is calculated to be one of the most efficiently activated by this symmetry breaking. Their early appearance and the $\sim$3~ps dielectric extrema provide experimental signatures of the structural reorganization, indicating a modified lattice potential in the photoexcited state (Supplementary \ref{si:phonon}). Beyond the electrodynamic inversion, the negative $\Delta\varepsilon_{\mathrm{im}}$ amplitude of the 2.7 THz mode is $\sim$3 times that of the 4.5 THz mode. Although the 2.7 THz mode has a lower static density of states (fig. \ref{fig:figS4.png}\textbf{B}), its threefold larger dielectric response indicates that the nonequilibrium dynamics are governed by mode-specific dipole coupling rather than available phase space.

Of the resolved phonon responses, the 2.7 and 8.0 THz modes constitute the most direct spectroscopic indicators of exciton localization. Because the optical activity of the 2.7 THz mode requires the C-centering translation to be broken at the excitation site, its emergence is intrinsically linked to exciton self-trapping. Its $\Delta\varepsilon_{\mathrm{im}}$ follows the sign inversion of the low-frequency background and builds up over the same $\sim$3 ps window as the localized dielectric state (Fig.~\ref{fig:fig3.png}\textbf{C}). Moreover, in contrast to the 5.7 THz mode, the 8.0 THz Ni--S stretching mode is suppressed during exciton formation and remains renormalized thereafter, reflecting a persistent local distortion of NiS$_6$ octahedra. Furthermore, the combined quantum energy of the two modes ($\sim$44 meV) lies within the uncertainty of $E_b$ ($\approx45\pm2$ meV), consistent with a multi-phonon contribution to self-trapping, as discussed below.

\section*{Discussion}\label{sec3}

The present measurements reveal a nonequilibrium pathway through which photoexcited charge-transfer carriers in NiPS$_3$ evolve from an initially itinerant state into a localized self-trapped exciton. As illustrated by the configuration coordinate picture in Fig. \ref{fig:fig4.png}, excitation across the ligand-to-metal charge-transfer (LMCT) gap generates high-energy carriers that rapidly bind into excitons and subsequently localize, accompanied by substantial lattice renormalization. The localization barrier $E_b\approx45\pm2$ meV is consistent with an elastic barrier separating delocalized and self-trapped excitonic configurations within the Rashba--Toyozawa framework of exciton self-trapping \cite{ExcitonInPhonon}. Taken together, the OPTP, TAS, and PL measurements indicate that charge-transfer excitations do not simply relax through a conventional energy dissipation pathway, but instead funnel into a self-trapped excitonic configuration stabilized by strong exciton-phonon coupling.

The central mechanistic implication of these observations is a cooperative two-phonon contribution of the 2.7 and 8.0 THz modes to exciton self-trapping. In the photoinduced transient state, both modes transform as odd-parity $A_u$, such that their linear coupling to a parity-definite free exciton vanishes ($\langle \psi | Q_{A_u} | \psi \rangle = 0 $) and neither alone can produce the symmetric deformation that defines a localization potential. By contrast, their direct product ($A_u \otimes A_u = A_g$) yields the quadratic term $\langle \psi | Q_{A_u} Q_{A_u} | \psi \rangle$, which describes a cooperative distortion compatible with the lattice relaxation that stabilizes self-trapped excitonic states. Because this quadratic term combines the zone-boundary 2.7 THz mode with the zone-center 8.0 THz mode, it conserves crystal momentum when localization removes the C-centering translation. This is reflected in the measured dynamics, in which the 8.0 THz mode is renormalized during exciton formation, whereas the 2.7 THz mode appears alongside the negative dielectric response that marks localization. Their combined quantum energy also falls within the uncertainty of $E_b$ and is compatible with this cooperative coupling. Although the $2\times5.7$ THz overtone ($B_u \otimes B_u = A_g$) satisfies the same symmetry and energy criteria, the prompt kinetic decoupling and steady amplitude (Fig.~\ref{fig:fig3.png}\textbf{C}) point to its non-dominant participation (Supplementary \ref{si:spectro} and \ref{si:srSTE}). Ultimately, the convergence of energetic, dynamical and parity constraints supports a cooperative two-phonon mechanism underlying self-trapped exciton formation in NiPS$_3$.

The present results provide a unified framework for understanding the diverse nonequilibrium states reported previously in NiPS$_3$. Earlier ultrafast studies primarily employed resonant or near-resonant excitation within the low-energy excitonic manifold, where photoexcitation directly addresses preexisting Zhang--Rice-like excitons and deposits little excess energy into the lattice \cite{RN3556,negsig}. Under these conditions, transient metallic behavior, population inversion, and long-lived nonequilibrium states have been reported. Excitation at 1.91 eV, by contrast, drives the inter-site $S~3p \rightarrow Ni~3d$ charge-transfer transition, generating high-energy unbound carriers and producing substantial charge perturbation of the local Ni--S bonding network. This excess energy enables activation of lattice coordinates that are not efficiently accessed under resonant excitonic pumping and promotes rapid carrier localization into self-trapped excitons. These observations indicate that the contrasting non-adiabatic evolution of NiPS$_3$ arises not from conflicting interpretations of the same excitation regime, but from fundamentally different photoexcitation pathways that access distinct regions of the coupled charge-lattice energy landscape.

A key enabling aspect of this work is the use of ultrabroadband terahertz spectroscopy. Extending the probe bandwidth to 15 THz provides simultaneous access to the emergent 2.7 THz vibrational modes and the intrinsic 8.0 THz Ni--S optical phonons that are involved in energy dissipation and lattice relaxation. This capability enables direct spectro-temporal tracking of the lattice coordinates involved in exciton localization and reveals correlations that would be inaccessible using conventional narrow-band THz probes. More broadly, the results highlight the importance of mode-selective exciton-phonon coupling in correlated van der Waals materials. The identification of specific lattice coordinates that mediate excitonic localization opens new opportunities to steer carrier self-trapping and manipulate emergent quantum states in low-dimensional materials through selective vibrational excitation with intense terahertz or mid-infrared fields.

\section*{Materials and Methods}\label{sec11}
\subsection*{Sample synthesis and characterization}\label{subsec4}

Single crystals of NiPS$_3$  were synthesized by chemical vapor transport (CVT). Stoichiometric amounts of Ni, P, and S powders (molar ratio 1:1:3) together with 0.15 g of I$_2$ transport agent were sealed in an evacuated quartz ampoule. Crystal growth was carried out in a two-zone furnace with source and growth-zone temperatures of 973 K and 923 K, respectively, for approximately two weeks. The furnace was subsequently cooled to room temperature at a rate of 100 K h$^{-1}$. 
For optical measurements, thin platelet-like crystals with typical dimensions of $2 \times 2 \times 0.09$ mm$^3$ were mechanically exfoliated from the bulk material to obtain optically flat surfaces. Magnetic characterization was performed using a Quantum Design Magnetic Property Measurement System (MPMS-5T). The Néel temperature was determined from the maximum in $d\chi/dT$, yielding $T_\mathrm{N}\approx155$ K, consistent with previous reports for high-quality NiPS$_3$ single crystals. Crystal structure and phase purity were further confirmed by X-ray diffraction (Supplementary fig. \ref{fig:figS1.png}).

\subsection*{Ultrabroadband optical-pump terahertz-probe spectroscopy}\label{subsec5}

Time-resolved optical-pump terahertz-probe (OPTP) measurements were performed using an air-plasma-based terahertz spectrometer. Ultrabroadband THz probe pulses were generated by two-color laser-induced plasma filamentation in nitrogen using 800 nm pulses ($\sim$40 fs) from a Ti:sapphire regenerative amplifier and their second harmonic generated in a $\beta$-barium borate (BBO) crystal. The THz electric fields were detected using air-biased coherent detection (ABCD), yielding phase-resolved measurements with spectral bandwidths exceeding 20 THz and an overall temporal resolution of better than 60 fs. The combination of ultrabroad spectral coverage and ultrafast temporal resolution enabled simultaneous tracking of the transient low-frequency vibrational modes together with the intrinsic Ni--S optical phonons at 5.7 and 8.0 THz that participate in the localized dynamics discussed in the main text. Photoexcitation was provided by an optical parametric amplifier driven by the regenerative amplifier. Unless otherwise stated, the excitation pulse was fixed at 650 nm central wavelength (1.91 eV) to selectively drive the ligand-to-metal charge-transfer transition in NiPS$_3$.

All OPTP measurements were performed in reflection geometry (RL path in Fig.~\ref{fig:fig2.png}\textbf{A}) to preserve the full detection bandwidth by avoiding strong attenuation of the THz probe by the infrared-active phonons of NiPS$_3$ in transmission measurements. One-dimensional dynamics were recorded by scanning the pump--probe delay while fixing the THz sampling position at the peak of the reflected THz waveform. Two-dimensional spectro-temporal measurements were obtained by independently scanning both the pump--probe and THz sampling delays, enabling extraction of the transient complex dielectric function. Temperature-dependent measurements were performed between 75 and 300 K using a continuous-flow cryostat (MicrostatHe, Oxford Instruments) equipped with CVD diamond windows to minimize THz absorption and dispersion across the measured spectral range.

\subsection*{Absorption, photoluminescence, and transient absorption spectroscopy}\label{subsec6}

\textbf{Optical absorption spectroscopy.} Steady-state optical spectra were measured using a UV--Vis--NIR spectrophotometer (Cary 5000, Varian) over the wavelength range of 200--1500 nm. Measurements were performed in diffuse-reflectance geometry to enhance sensitivity to weak optical transitions, including the parity-forbidden crystal-field ($d$--$d$) excitations of Ni$^{2+}$.

\noindent\textbf{Photoluminescence spectroscopy.} Photoluminescence (PL) measurements were carried out using a 637 nm continuous-wave laser for excitation. The emitted PL was collected in a backscattering geometry, dispersed by a Czerny--Turner spectrometer (SR500i, Andor), and detected using a cooled InGaAs array detector (DU491A-1.7, Andor).

\noindent\textbf{Transient absorption spectroscopy.} Room-temperature transient absorption spectroscopy (TAS) measurements were performed using femtosecond pump pulses centered at 592 nm (2.09 eV) with a fluence of 1.5 mJ cm$^{-2}$. The excitation conditions were chosen to generate a carrier density comparable to that used in the OPTP experiments, enabling direct correlation between the excited-state population dynamics measured by TAS and the transient electrodynamic response obtained from terahertz spectroscopy.

%%%%%%%%%%%%%%%% MAIN TEXT FIGURES %%%%%%%%%%%%%%%
% SA: figures go at the end of the main text, not in the middle of it.

\begin{figure} % Do NOT use \begin{figure*}
\centering
\includegraphics[width=0.95\linewidth]{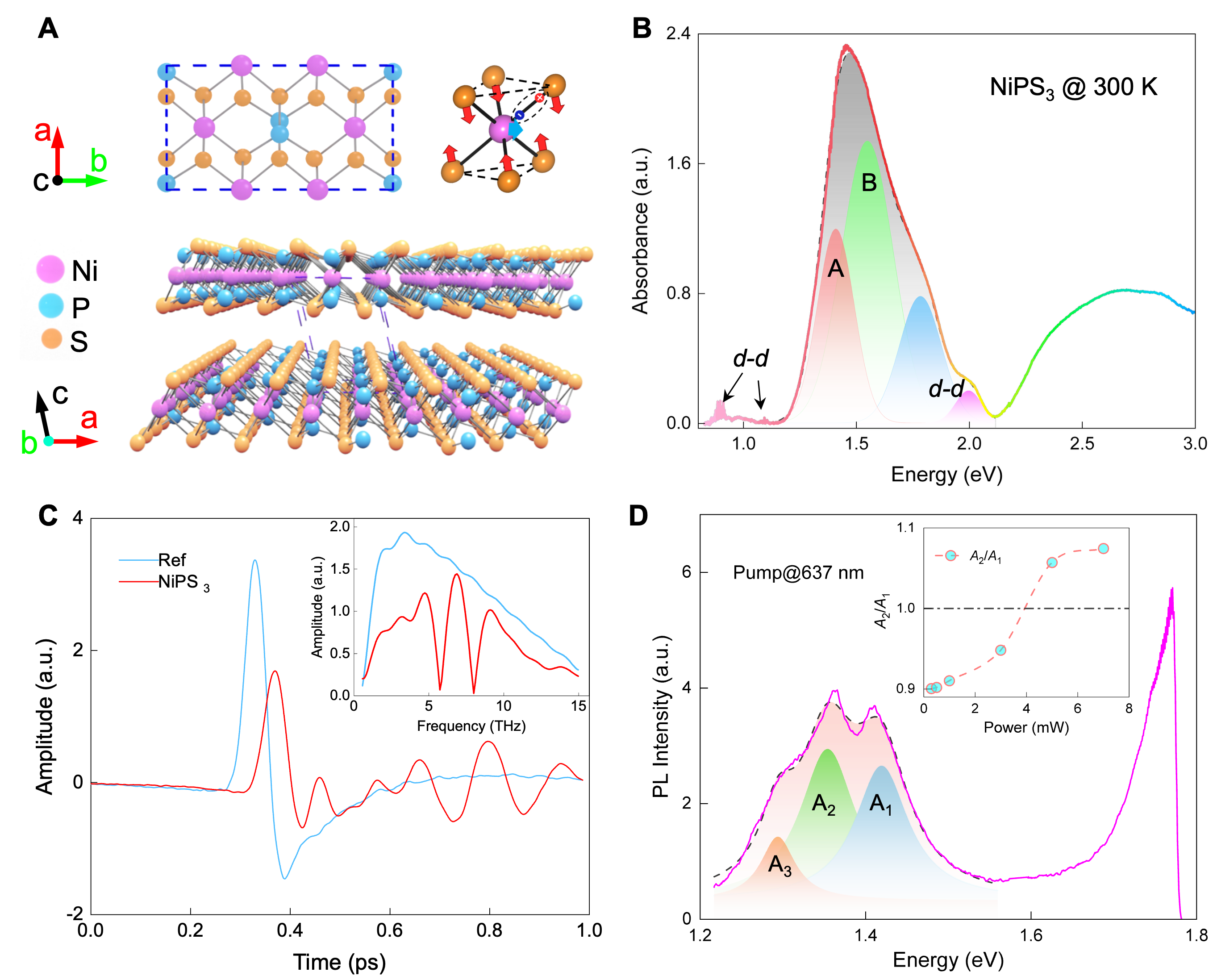}
\caption{\textbf{Crystal structure, optical transitions, and vibrational properties of NiPS$_3$.} (\textbf{A}) Crystal structure of NiPS$_3$ in top and side views together with the local NiS$_6$ octahedral coordination (upper right). The schematic illustrates the ligand-to-metal charge-transfer excitation from S $3p$ to Ni $3d$ orbitals and the associated coupling to lattice vibrations. (\textbf{B}) Room-temperature optical absorbance spectrum measured in diffuse-reflectance geometry. Gaussian decomposition resolves the excitonic resonance near 1.4 eV (A, red), the LMCT absorption at $\sim$1.6 eV (B, green), and higher-energy crystal-field ($d-d$ ) excitations near 1.78 and 2.0 eV (blue and pink). (\textbf{C}) Time-domain terahertz electric-field waveforms transmitted through free space (cyan) and a NiPS$_3$ crystal (red). Inset: Corresponding frequency-domain spectra, featuring infrared-active optical phonons at $\sim$5.7 and 8.0 THz. (\textbf{D}) Room-temperature photoluminescence (PL) spectrum under continuous-wave 637 nm excitation. Shaded areas: Gaussian components of the dominant excitonic emissions (A$_1$ and A$_2$) and a lower-energy sideband (A$_3$) that appears at excitation power above 11 kW/cm$^2$. Inset: excitation-power dependence of the integrated intensity ratio A$_2$/A$_1$.}

\label{fig:fig1.png}
\end{figure}

\begin{figure} % Do NOT use \begin{figure*}
\centering
\includegraphics[width=\linewidth]{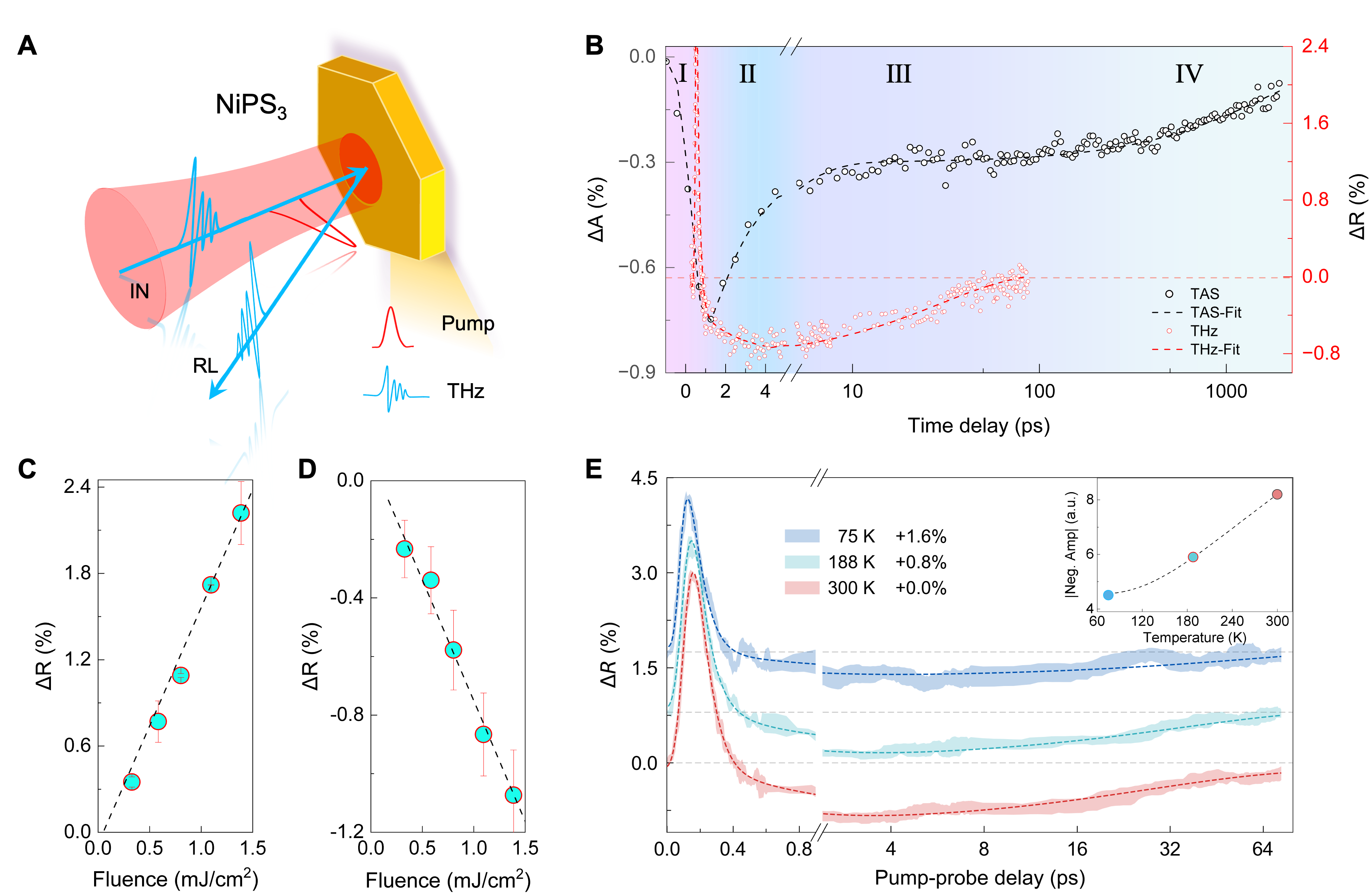}
\caption{\textbf{Ultrafast carrier dynamics following charge-transfer excitation in NiPS$_3$.} (\textbf{A}) Schematic of the ultrabroadband optical-pump terahertz-probe (OPTP) setup in reflection (RL) geometry. (\textbf{B}) Room-temperature transient response measured by OPTP (red circles) and transient absorption spectroscopy (TAS, black circles). Dashed curves represent convolved multi-exponential fits. Regions I--IV denote the successive dynamical stages associated with carrier generation, ultrafast exciton formation, exciton localization, and subsequent relaxation. (\textbf{C}~and~\textbf{D}) Pump-fluence dependence of the OPTP signal extrema corresponding to the initial positive response (\textbf{C}) and the subsequent negative response (\textbf{D}). Dashed lines are linear fits. Error bars represent standard deviations from repeated measurements. (\textbf{E}) OPTP dynamics measured at 75, 188, and 300 K. Shaded bands denote the data with their noise envelope, colored dashed lines are exponential fits, and grey dashed lines mark the zero level of each trace. Traces are vertically offset by the values indicated for clarity. Inset: Arrhenius analysis (dashed line) of the temperature-dependent absolute amplitude of the negative response.}

\label{fig:fig2.png}
\end{figure}

\begin{figure} % Do NOT use \begin{figure*}
\centering
\includegraphics[width=\linewidth]{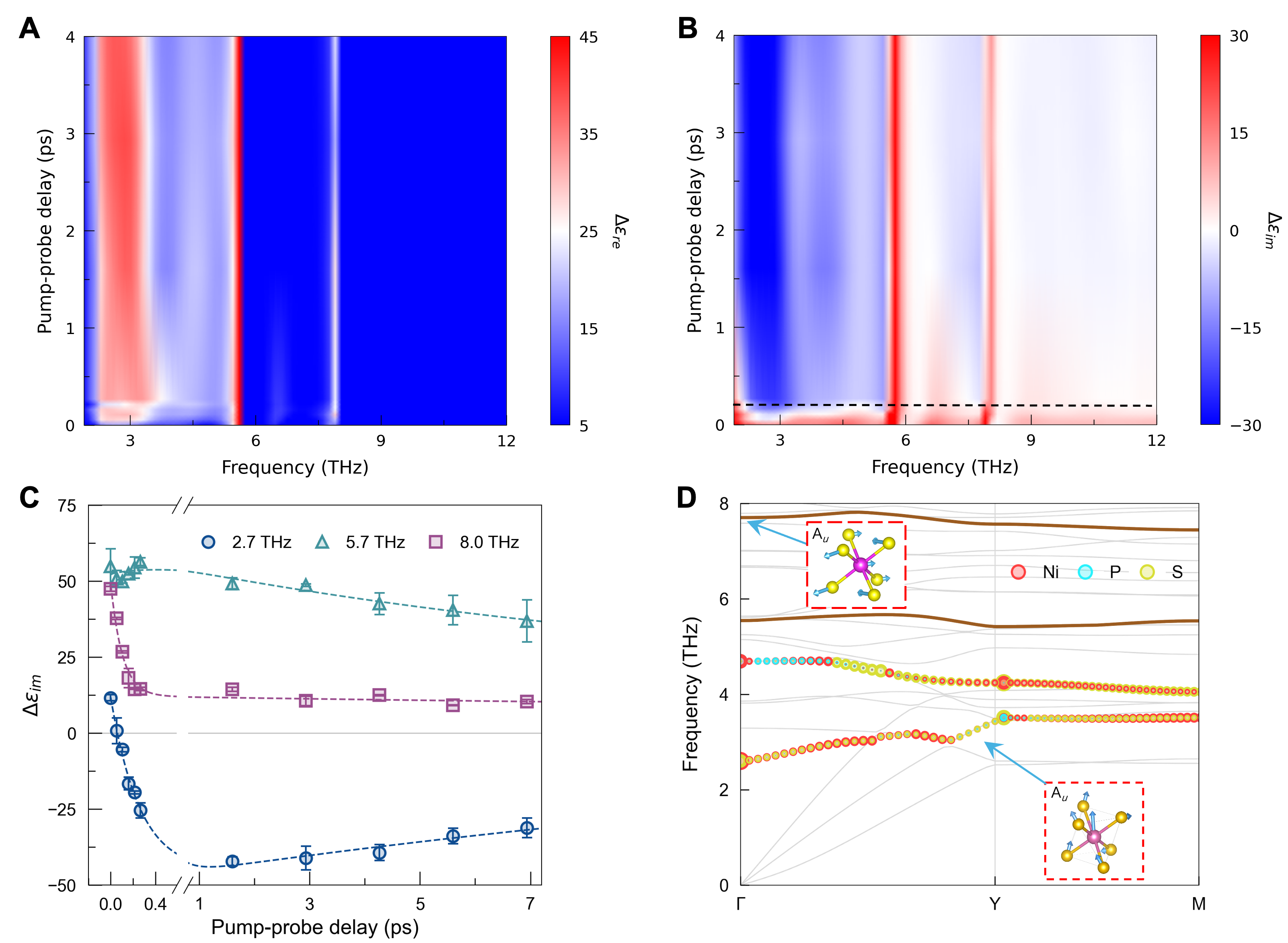}
\caption{\textbf{Spectro-temporal mapping of transient dielectric and vibrational dynamics in NiPS$_3$.}
(\textbf{A}~and~\textbf{B}) Two-dimensional spectro-temporal maps of the pump-induced changes in the real ($\Delta\varepsilon_{\mathrm{re}}$, \textbf{A}) and imaginary ($\Delta\varepsilon_{\mathrm{im}}$, \textbf{B}) components of the dielectric function at 300 K following 1.91 eV excitation.
(\textbf{C}) Temporal evolution of the imaginary dielectric response($\Delta\varepsilon_{im}$) at 2.7~THz (blue circles), 5.7 THz (green triangles), 8.0~THz (red squares). Dashed curves are exponential fits, and error bars denote standard deviations. 
(\textbf{D}) Calculated equilibrium phonon dispersion of NiPS$_3$ (solid lines) overlaid with the emergent transient branches (bubbles) derived from the charge-transfer excited state. Dark red denotes the two intrinsically THz-active modes; grey, the remainder. Symbol colors indicate the fractional atomic contributions from Ni (red), P (cyan), and S (yellow). Insets show the corresponding real-space vibrational eigenvectors of the 8.0~THz and 2.7 THz A$_u$ modes.}

\label{fig:fig3.png}
\end{figure}

\begin{figure} % Do NOT use \begin{figure*}
\centering
\includegraphics[width=0.9\linewidth]{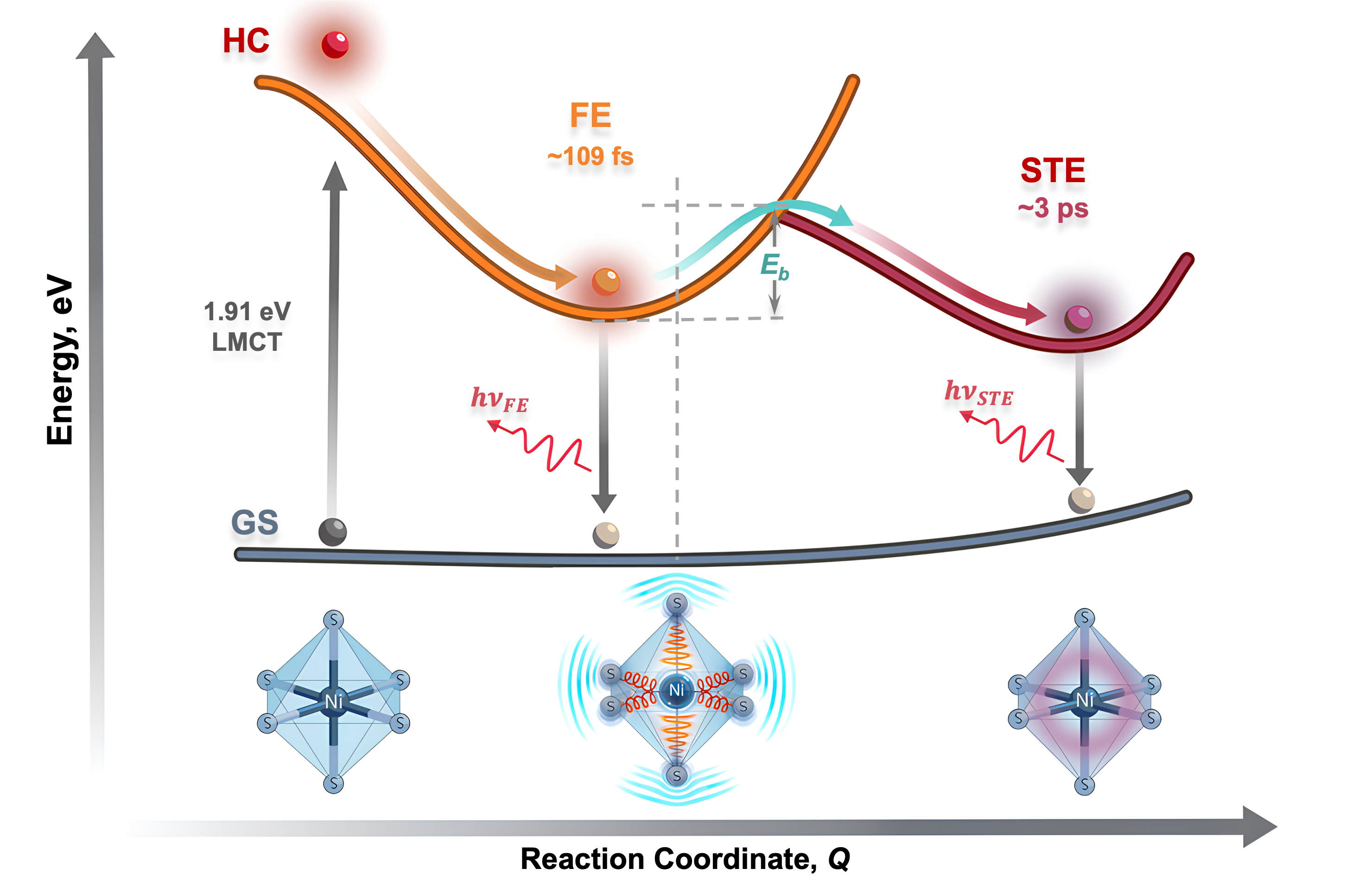}
\caption{\textbf{Configuration-coordinate model for exciton self-trapping in NiPS$_3$.}
Schematic potential-energy surfaces along the generalized Ni--S lattice displacement coordinate $Q$, showing the correlated ground state (GS, dark grey), charge-transfer hot carriers (HC, red), free exciton (FE, orange), and self-trapped exciton (STE, dark red). Vertical arrows denote the initial 1.91 eV excitation and the eventual radiative recombination. The extracted activation barrier $E_b \approx 45\pm2$ meV sets the kinetic threshold for spatial localization. The wavy red arrows indicate the dual radiative recombination pathways originating from both the free ($h\nu_{FE}$) and self-trapped ($h\nu_{STE}$) excitons. The bottom panels schematically illustrate the microscopic structural evolution, tracking the progressive distortion of the local NiS$_6$ octahedra during the polaronic collapse.
}

\label{fig:fig4.png}
\end{figure}

%%%%%%%%%%%%%%%% REFERENCES %%%%%%%%%%%%%%%

\clearpage % Clear all remaining figures and tables then start a new page

\bibliography{sn-bibliography}

@article{CTgap300K,
  title={Pressure-induced enhancement and retainability of optoelectronic properties of {NiPS$_3$}},
  author={Fang, Sixue and Li, Quanjun and Li, Zonglun and Dong, Qing and Jing, Xiaoling and Li, Chenyi and Li, Haiyan and Liu, Bo and Liu, Ran and Liu, Bingbing},
  journal={Mater. Res. Lett.},
  volume={11},
  number={2},
  pages={134--142},
  year={2023},
  publisher={Taylor \& Francis}
}

@article{2Dmagnets1,
  title={Layer-dependent ferromagnetism in a {van der Waals} crystal down to the monolayer limit},
  author={Huang, Bevin and Clark, Genevieve and Navarro-Moratalla, Efr{\'e}n and Klein, Dahlia R and Cheng, Ran and Seyler, Kyle L and Zhong, Ding and Schmidgall, Emma and McGuire, Michael A and Cobden, David H and Yao, Wang and Xiao, Di and Jarillo-Herrero, Pablo and Xu, Xiaodong},
  journal={Nature},
  volume={546},
  number={7657},
  pages={270--273},
  year={2017},
  publisher={Nature Publishing Group UK London}
}

@article{2Dmagnets2,
  title={Exciton-coupled coherent magnons in a {2D} semiconductor},
  author={Bae, Youn Jue and Wang, Jue and Scheie, Allen and Xu, Junwen and Chica, Daniel G and Diederich, Geoffrey M and Cenker, John and Ziebel, Michael E and Bai, Yusong and Ren, Haowen and Dean, Cory R. and Delor, Milan and Xu, Xiaodong and Roy, Xavier and Kent, Andrew D. and Zhu, Xiaoyang},
  journal={Nature},
  volume={609},
  number={7926},
  pages={282--286},
  year={2022},
  publisher={Nature Publishing Group UK London}
}

@article{2Dmagnets3,
  title={Magnetism in two-dimensional {van der Waals} materials},
  author={Burch, Kenneth S and Mandrus, David and Park, Je-Geun},
  journal={Nature},
  volume={563},
  number={7729},
  pages={47--52},
  year={2018},
  publisher={Nature Publishing Group UK London}
}

@article{abinitio,
  title={An ab initio study of electron-hole pairs in a correlated {van der Waals} antiferromagnet: {NiPS$_3$}},
  author={Lane, Christopher and Zhu, Jian-Xin},
  year={2022},
  note={arXiv:2209.13051}
}

@article{polaron,
   author = {Franchini, Cesare and Reticcioli, Michele and Setvin, Martin and Diebold, Ulrike},
   title = {Polarons in materials},
   journal = {Nat. Rev. Mater.},
   volume = {6},
   number = {7},
   pages = {560-586},
   ISSN = {2058-8437},
   OPTdoi = {10.1038/s41578-021-00289-w},
   year = {2021},
   type = {Journal Article}
}

@article{excitonControl,
  title={Exciton control in a room temperature bulk semiconductor with coherent strain pulses},
  author={Baldini, Edoardo and Dominguez, Adriel and Palmieri, Tania and Cannelli, Oliviero and Rubio, Angel and Ruello, Pascal and Chergui, Majed},
  journal={Sci. Adv.},
  volume={5},
  number={11},
  pages={eaax2937},
  year={2019},
  publisher={American Association for the Advancement of Science}
}

@article{manybody,
   author = {Pei, Jiajie and Yang, Jiong and Yildirim, Tanju and Zhang, Han and Lu, Yuerui},
   title = {Many-Body Complexes in {2D} Semiconductors},
   journal = {Adv. Mater.},
   volume = {31},
   number = {2},
   pages = {1706945},
   ISSN = {0935-9648},
   OPTdoi = {10.1002/adma.201706945},
   xurl = {https://doi.org/10.1002/adma.201706945},
   year = {2019},
   type = {Journal Article}
}

@article{ddtransition3,
   author = {Afanasiev, Dmytro and Hortensius, Jorrit R. and Matthiesen, Mattias and Mañas-Valero, Samuel and Šiškins, Makars and Lee, Martin and Lesne, Edouard and van der Zant, Herre S. J. and Steeneken, Peter G. and Ivanov, Boris A. and Coronado, Eugenio and Caviglia, Andrea D.},
   title = {Controlling the anisotropy of a {van der Waals} antiferromagnet with light},
   journal = {Sci. Adv.},
   volume = {7},
   number = {23},
   pages = {eabf3096},
   OPTdoi = {10.1126/sciadv.abf3096},
   xurl = {http://advances.sciencemag.org/content/7/23/eabf3096.abstract
https://www.ncbi.nlm.nih.gov/pmc/articles/PMC8172129/pdf/abf3096.pdf},
   year = {2021},
   type = {Journal Article}
}

@article{bandgap_300K,
author = {Wang, Fengmei and Shifa, Tofik A. and Yu, Peng and He, Peng and Liu, Yang and Wang, Feng and Wang, Zhenxing and Zhan, Xueying and Lou, Xiaoding and Xia, Fan and He, Jun},
title = {New Frontiers on {van der Waals} Layered Metal Phosphorous Trichalcogenides},
journal = {Adv. Funct. Mater.},
volume = {28},
number = {37},
pages = {1802151},
OPTdoi = {10.1002/adfm.201802151},
xurl = {https://advanced.onlinelibrary.wiley.com/doi/abs/10.1002/adfm.201802151},
eprint = {https://advanced.onlinelibrary.wiley.com/doi/pdf/10.1002/adfm.201802151},
year = {2018}
}

@incollection{ExcitonInPhonon,
author="Ueta, Masayasu
and Kanzaki, Hiroshi
and Kobayashi, Koichi
and Toyozawa, Yutaka
and Hanamura, Eiichi",
title="Theory of Excitons in Phonon Fields",
booktitle="Excitonic Processes in Solids",
year="1986",
publisher="Springer Berlin Heidelberg",
address="Berlin, Heidelberg",
pages="203--284",
isbn="978-3-642-82602-3",
OPTdoi="10.1007/978-3-642-82602-3_4",
  series={Springer Series in Solid-State Sciences},
  volume={60}
}

@article{RN3556,
   author = {Belvin, C. A. and Baldini, E. and Ozel, I. O. and Mao, D. and Po, H. C. and Allington, C. J. and Son, S. and Kim, B. H. and Kim, J. and Hwang, I. and Kim, J. H. and Park, J. G. and Senthil, T. and Gedik, N.},
   title = {Exciton-driven antiferromagnetic metal in a correlated {van der Waals} insulator},
   journal = {Nat. Commun.},
   volume = {12},
   number = {1},
   pages = {4837},
   ISSN = {2041-1723 (Electronic)
2041-1723 (Linking)},
   OPTdoi = {10.1038/s41467-021-25164-8},
   xurl = {https://www.ncbi.nlm.nih.gov/pubmed/34376692},
   year = {2021},
   type = {Journal Article}
}

@ARTICLE{Kim:RN3583,
   author = {Kim, S. Y. and Kim, T. Y. and Sandilands, L. J. and Sinn, S. and Lee, M. C. and Son, J. and Lee, S. and Choi, K. Y. and Kim, W. and Park, B. G. and Jeon, C. and Kim, H. D. and Park, C. H. and Park, J. G. and Moon, S. J. and Noh, T. W.},
   title = {Charge-Spin Correlation in {van der Waals} Antiferromagnet {NiPS$_3$}},
   journal = {Phys. Rev. Lett.},
   volume = {120},
   number = {13},
   pages = {136402},
   ISSN = {1079-7114 (Electronic) 0031-9007 (Linking)},
   OPTdoi = {10.1103/PhysRevLett.120.136402},
   year = {2018},
   type = {Journal Article}
}

@article{RN3592,
   author = {Pistawala, Nashra and Kumar, Ankit and Negi, Devesh and Rout, Dibyata and Harnagea, Luminita and Saha, Surajit and Singh, Surjeet},
   title = {Probing electron–phonon coupling in magnetic {van der Waals} material {NiPS$_3$}: A non-magnetic site-dilution study},
   journal = {2D Mater.},
   volume = {11},
   number = {2},
   pages = {025035},
   ISSN = {2053-1583},
   OPTdoi = {10.1088/2053-1583/ad2f44},
   year = {2024},
   type = {Journal Article}
}

@article{RN3589,
   author = {Klaproth, T. and Aswartham, S. and Shemerliuk, Y. and Selter, S. and Janson, O. and van den Brink, J. and B{\"u}chner, B. and Knupfer, M. and Pazek, S. and Mikhailova, D. and Efimenko, A. and Hayn, R. and Savoyant, A. and Gubanov, V. and Koitzsch, A.},
   title = {Origin of the Magnetic Exciton in the {van der Waals} Antiferromagnet {NiPS$_3$}},
   journal = {Phys. Rev. Lett.},
   volume = {131},
   number = {25},
   pages = {256504},
   ISSN = {1079-7114 (Electronic)
0031-9007 (Linking)},
   OPTdoi = {10.1103/PhysRevLett.131.256504},
   xurl = {https://www.ncbi.nlm.nih.gov/pubmed/38181357},
   year = {2023},
   type = {Journal Article}
}

@article{vasp1,
  title = {Efficient iterative schemes for ab initio total-energy calculations using a plane-wave basis set},
  author = {Kresse, G. and Furthm\"uller, J.},
  journal = {Phys. Rev. B},
  volume = {54},
  issue = {16},
  pages = {11169--11186},
  numpages = {0},
  year = {1996},
  month = {Oct},
  publisher = {American Physical Society},
  OPTdoi = {10.1103/PhysRevB.54.11169},
  xurl = {https://link.aps.org/doi/10.1103/PhysRevB.54.11169}
}

@article{NPS_Struc,
   author = {Lane, Christopher and Zhu, Jian-Xin},
   title = {Thickness dependence of electronic structure and optical properties of a correlated {van der Waals} antiferromagnetic {NiPS$_3$} thin film},
   journal = {Phys. Rev. B},
   volume = {102},
   number = {7},
   ISSN = {2469-9950},
   OPTdoi = {10.1103/PhysRevB.102.075124},
   year = {2020},
   type = {Journal Article},
  pages={075124}
}

@article{metaGGA,
  title = {Strongly Constrained and Appropriately Normed Semilocal Density Functional},
  author = {Sun, Jianwei and Ruzsinszky, Adrienn and Perdew, John P.},
  journal = {Phys. Rev. Lett.},
  volume = {115},
  issue = {3},
  pages = {036402},
  numpages = {6},
  year = {2015},
  month = {Jul},
  publisher = {American Physical Society},
  OPTdoi = {10.1103/PhysRevLett.115.036402},
  xurl = {https://link.aps.org/doi/10.1103/PhysRevLett.115.036402}
}

@article{ddtransition1,
   author = {Wang, Xingzhi and Cao, Jun and Li, Hua and Lu, Zhengguang and Cohen, Arielle and Haldar, Anubhab and Kitadai, Hikari and Tan, Qishuo and Burch, Kenneth S. and Smirnov, Dmitry and Xu, Weigao and Sharifzadeh, Sahar and Liang, Liangbo and Ling, Xi},
   title = {Electronic Raman scattering in the {2D} antiferromagnet {NiPS$_3$}},
   journal = {Sci. Adv.},
   volume = {8},
   number = {2},
   pages = {eabl7707},
   year = {2022},
   OPTdoi = {10.1126/sciadv.abl7707},
   xurl = {https://doi.org/10.1126/sciadv.abl7707},
   type = {Journal Article}
}

@article{phonons,
   author = {Hashemi, Arsalan and Komsa, Hannu-Pekka and Puska, Martti and Krasheninnikov, Arkady V.},
   title = {Vibrational Properties of Metal Phosphorus Trichalcogenides from First-Principles Calculations},
   journal = {J. Phys. Chem. C},
   volume = {121},
   number = {48},
   pages = {27207-27217},
   ISSN = {1932-7447},
   OPTdoi = {10.1021/acs.jpcc.7b09634},
   xurl = {https://doi.org/10.1021/acs.jpcc.7b09634},
   year = {2017},
   type = {Journal Article}
}

@article{exciton,
   author = {Kang, S. and Kim, K. and Kim, B. H. and Kim, J. and Sim, K. I. and Lee, J. U. and Lee, S. and Park, K. and Yun, S. and Kim, T. and Nag, A. and Walters, A. and Garcia-Fernandez, M. and Li, J. and Chapon, L. and Zhou, K. J. and Son, Y. W. and Kim, J. H. and Cheong, H. and Park, J. G.},
   title = {Coherent many-body exciton in {van der Waals} antiferromagnet {NiPS$_3$}},
   journal = {Nature},
   volume = {583},
   number = {7818},
   pages = {785-789},
   ISSN = {1476-4687 (Electronic)
0028-0836 (Linking)},
   OPTdoi = {10.1038/s41586-020-2520-5},
   xurl = {https://www.ncbi.nlm.nih.gov/pubmed/32690938},
   year = {2020},
   type = {Journal Article}
}

@article{excitonphonon,
   author = {Hwangbo, Kyle and Zhang, Qi and Jiang, Qianni and Wang, Yong and Fonseca, Jordan and Wang, Chong and Diederich, Geoffrey M. and Gamelin, Daniel R. and Xiao, Di and Chu, Jiun-Haw and Yao, Wang and Xu, Xiaodong},
   title = {Highly anisotropic excitons and multiple phonon bound states in a {van der Waals} antiferromagnetic insulator},
   journal = {Nat. Nanotechnol.},
   volume = {16},
   number = {6},
   pages = {655-660},
   ISSN = {1748-3395},
   OPTdoi = {10.1038/s41565-021-00873-9},
   xurl = {https://doi.org/10.1038/s41565-021-00873-9},
   year = {2021},
   type = {Journal Article}
}

@article{excitonlife,
   author = {Li, Yuanhe and Liang, Gaoming and Kong, Chongtao and Sun, Baoquan and Zhang, Xinhui},
   title = {Charge-Transfer-Mediated Exciton Dynamics in {van der Waals} Antiferromagnet {NiPS$_3$}},
   journal = {Adv. Funct. Mater.},
   volume = {34},
   number = {40},
   pages = {2402161},
   ISSN = {1616-301X},
   OPTdoi = {10.1002/adfm.202402161},
   xurl = {https://onlinelibrary.wiley.com/doi/abs/10.1002/adfm.202402161},
    year = {2024},
   type = {Journal Article}
}

@article{phonopy,
title = {First principles phonon calculations in materials science},
journal = {Scr. Mater.},
volume = {108},
pages = {1-5},
year = {2015},
issn = {1359-6462},
OPTdoi = {10.1016/j.scriptamat.2015.07.021},
xurl = {https://www.sciencedirect.com/science/article/pii/S1359646215003127},
author = {Atsushi Togo and Isao Tanaka}
}

@article{ir_active,
  title={Interlayer interaction and Davydov splitting in antiferromagnetic few-layer {NiPS$_3$}},
  author={Nguyen, Manh Hong and Park, Giung and Park, Je-Geun and Cheong, Hyeonsik},
  journal={npj Quantum Mater.},
  volume={10},
  number={1},
  pages={65},
  year={2025},
  publisher={Nature Publishing Group UK London}
}

@article{ir_active1,
  title={Anisotropic phonon and magnon vibration and gate-tunable optoelectronic properties of nickel thiophosphite},
  author={Muhammad, Zahir and Szpakowski, Jan and Abbas, Ghulam and Zu, Lin and Islam, Rajibul and Wang, Yan and Wali, Faiz and Karmakar, Arka and Molas, Maciej R and Zhang, Yue and Zhu, Ling and Zhao, Weisheng and Zhang, Han},
  journal={2D Mater.},
  volume={10},
  number={2},
  pages={025001},
  year={2023},
  publisher={IOP Publishing}
}

@article{MoS2_exciton_timescale,
  title={The ultrafast onset of exciton formation in {2D} semiconductors},
  author={Trovatello, Chiara and Katsch, Florian and Borys, Nicholas J and Selig, Malte and Yao, Kaiyuan and Borrego-Varillas, Rocio and Scotognella, Francesco and Kriegel, Ilka and Yan, Aiming and Zettl, Alex and Schuck, P. James and Knorr, Andreas and Cerullo, Giulio and Dal Conte, Stefano},
  journal={Nat. Commun.},
  volume={11},
  number={1},
  pages={5277},
  year={2020},
  publisher={Nature Publishing Group UK London}
}

@article{WSe2_exciton_timescale,
  title={Direct observation of ultrafast exciton formation in a monolayer of {WSe$_2$}},
  author={Steinleitner, Philipp and Merkl, Philipp and Nagler, Philipp and Mornhinweg, Joshua and Sch{\"u}ller, Christian and Korn, Tobias and Chernikov, Alexey and Huber, Rupert},
  journal={Nano Lett.},
  volume={17},
  number={3},
  pages={1455--1460},
  year={2017},
  publisher={ACS Publications}
}

@article{negsig,
  title={Long-lived population inversion in resonantly driven excitonic antiferromagnet},
  author={Warshauer, Jacob A and Chen, Huyongqing and Bustamante Lopez, Daniel Alejandro and Tan, Qishuo and Tang, Jing and Ling, Xi and Hu, Wanzheng},
  journal={Phys. Rev. Lett.},
  volume={134},
  number={1},
  pages={016901},
  year={2025},
  publisher={APS}
}

@article{phyNPS,
    title = {Optical transitions, {XPS}, electronic states in {NiPS$_3$}},
    volume = {65},
    issn = {0301-0104},
    OPTdoi = {10.1016/0301-0104(82)85205-1},
    number = {3},
    journal = {Chem. Phys.},
    author = {Piacentini, M. and Khumalo, F. S. and Olson, C. G. and Anderegg, J. W. and Lynch, D. W.},
    month = mar,
    year = {1982},
    pages = {289--304},
}

@article{elecphonon,
    title = {Magnetically brightened dark electron-phonon bound states in a van der {Waals} antiferromagnet},
    volume = {13},
    issn = {2041-1723 (Electronic) 2041-1723 (Linking)},
    OPTdoi = {10.1038/s41467-021-27741-3},
    number = {1},
    journal = {Nat. Commun.},
    author = {Erge{\c{c}}en, E. and Ilyas, B. and Mao, D. and Po, H. C. and Yilmaz, M. B. and Kim, J. and Park, J. G. and Senthil, T. and Gedik, N.},
    month = jan,
    year = {2022},
    pmcid = {PMC8748959},
    pages = {98},
}

@article{pgfm,
author = {Stock, Daniel and Wigger, Anna K. and Haring Bol{\'\i}var, Peter},
journal = {J. Opt. Soc. Am. B},
number = {3},
pages = {815--824},
publisher = {Optica Publishing Group},
title = {Reducing errors in {THz} material parameter determination by model-based time-domain extraction methods},
volume = {38},
month = {Mar},
year = {2021},
xurl = {https://opg.optica.org/josab/abstract.cfm?URI=josab-38-3-815},
OPTdoi = {10.1364/JOSAB.415144},
}

@article{PTMM,
  title = {Intrinsic complications in the analysis of optical-pump, terahertz probe experiments},
  author = {Nienhuys, Han-Kwang and Sundstr\"om, Villy},
  journal = {Phys. Rev. B},
  volume = {71},
  issue = {23},
  pages = {235110},
  numpages = {8},
  year = {2005},
  month = {Jun},
  publisher = {American Physical Society},
  OPTdoi = {10.1103/PhysRevB.71.235110},
  xurl = {https://link.aps.org/doi/10.1103/PhysRevB.71.235110}
}

@book{high_orderdd,
  title={Inorganic Electronic Spectroscopy},
  author={Lever, A. B. P.},
  lccn={84013490},
  series={Studies in physical and theoretical chemistry},
  xurl={https://books.google.dk/books?id=NRd1swEACAAJ},
  year={1984},
  publisher={Elsevier},
  edition={2nd},
  volume={33}
}

@article{enhance_dd_absorp_CT,
  title = {Optical-absorption spectra of the layered transition-metal thiophosphates {MPS$_3$ (M=Mn, Fe, and Ni)}},
  author = {Joy, P. A. and Vasudevan, S.},
  journal = {Phys. Rev. B},
  volume = {46},
  issue = {9},
  pages = {5134--5141},
  numpages = {0},
  year = {1992},
  month = {Sep},
  publisher = {American Physical Society},
  OPTdoi = {10.1103/PhysRevB.46.5134},
  xurl = {https://link.aps.org/doi/10.1103/PhysRevB.46.5134}
}

@book{CTmaterial_e-phon_couple,
  title={Polarons in advanced materials},
  volume={103},
  year={2007},
  publisher={Springer},
  editor={Alexandrov, Alexandre S.},
  series={Springer Series in Materials Science},
  address={Dordrecht}
}

@book{STEbook,
  title={Self-trapped excitons},
  author={Song, K. S. and Williams, Richard T},
  year={1996},
  publisher={Springer},
  edition={2nd},
  address={Berlin}
}

@book{MNBYBook,
  title={Many-particle physics},
  author={Mahan, Gerald D},
  year={2000},
  publisher={Kluwer Academic/Plenum Publishers},
  edition={3rd},
  address={New York}
}

@article{multiDegree,
  title={Towards properties on demand in quantum materials},
  author={Basov, D. N. and Averitt, R. D. and Hsieh, D.},
  journal={Nat. Mater.},
  volume={16},
  number={11},
  pages={1077--1088},
  year={2017},
  publisher={Nature Publishing Group UK London}
}

@article{lightDriveLattice,
  title={Engineering crystal structures with light},
  author={Disa, Ankit S and Nova, Tobia F and Cavalleri, Andrea},
  journal={Nat. Phys.},
  volume={17},
  number={10},
  pages={1087--1092},
  year={2021},
  publisher={Nature Publishing Group UK London}
}

@article{non-adiabatic-dynamics,
  title={Colloquium: Nonthermal pathways to ultrafast control in quantum materials},
  author={de la Torre, Alberto and Kennes, Dante M and Claassen, Martin and Gerber, Simon and McIver, James W and Sentef, Michael A},
  journal={Rev. Mod. Phys.},
  volume={93},
  number={4},
  pages={041002},
  year={2021},
  publisher={APS}
}
\bibliographystyle{sciencemag}

%%%%%%%%%%%%%%%% ACKNOWLEDGEMENTS %%%%%%%%%%%%%%%
%\section*{Acknowledgments}
%\paragraph*{Funding:}
%B.Z. acknowledge support from the \textcolor{red}{XXXX}
%\paragraph*{Author contributions:}
%L.C. and B.Z. conceived the study. L.C. and B.Z. designed and performed the experiments and analysed the data, supported by K.Z. Q.P. A.H.O. O.N. and X.H.,
%L.C. performed the DFT simulations, supported by K.Z. (S.M. Y.W. C.W.),
%X.L. Y.S. and Y.Z. grew the high-quality single crystal used in the experiments. Q.L. exfoliated the sample under the guidance of S.X.,
%L.C. B.Z. K.Z. and P.U.J. \textcolor{red}{XXX} interpreted the data.
%L.C. and B.Z. led the paper preparation with input from all the authors. B.Z. and P.U.J. supervised the project.
%The authors also thank the helpfull disscussion from H.F.
%\paragraph*{Competing interests:}
%The authors declare no competing interests.
%\paragraph*{Data, code and materials availability:}
% Filled in from overleaf-draft/sn-article-work-with-iop.tex (Overleaf commit bbe6f5c): its Acknowledgments
% paragraph (funding only) under Funding:, its Author contribution paragraph under Author contributions:.
\section*{Acknowledgments}
\paragraph*{Funding:}
This work was partially supported by the Independent Research Fund Denmark (Grant Nos. 2035-00365B and 6211-00182B), the Novo Nordisk Foundation (Grant No. NNF24OC0094858), and the Villum Fonden (VIL88455). X.H. and S.X. acknowledge support from Villum Fonden (VIL70202, VIL78289). X.L. and Y.S. acknowledge support from the National Key Research and Development Program of China (Grants No. 2021YFA1600200 and No. 2023YFA1607402) and the National Natural Science Foundation of China (Grant No. 12274412).
\paragraph*{Author contributions:}
L.C. and B.Z. conceived and coordinated the study. L.C. and B.Z. designed and performed the THz spectroscopy experiments and analyzed the experimental data, with assistance from A.H.O., O.N., and P.U.J.; G.Li and K.Z. performed the transient absorption spectroscopy measurements, with K.Z. also contributing to data analysis and optical absorption spectroscopy. Q.P. contributed to the optical absorption spectroscopy measurements. X.H. and S.X. performed sample exfoliation and photoluminescence measurements. L.C. led the density functional theory calculations, with substantial contributions to the theoretical modeling and analysis from S.M., C.W., and Y.W.; X.L., Y.S., and Y.Z. grew the NiPS$_3$ single crystals and performed the X-ray diffraction measurements. S.Z., G.Liao, and Y.L. contributed to scientific discussions, interpretation of the results, and critical review of the manuscript. L.C. and B.Z. prepared the initial manuscript, with input from all authors. B.Z. and P.U.J. supervised the project. All authors reviewed and approved the final manuscript.
\paragraph*{Competing interests:}
There are no competing interests to declare.
\paragraph*{Data, code and materials availability:}
All data needed to evaluate the conclusions in the paper are present in the paper and/or the Supplementary Materials.

%%%%%%%%%%%%%%%% SUPPLEMENT LIST %%%%%%%%%%%%%%%
% Refs. 32 onward are cited only in the supplement (31 references are cited in the main text).
\subsection*{Supplementary materials}
Supplementary Text\\
Figs. S1 to S8\\
References \textit{(32-\arabic{enumiv})} % automatically fills out the last reference number

%%%%%%%%%%%%%%%% END OF MAIN TEXT %%%%%%%%%%%%%%%

\newpage

%%%%%%%%%%%%%%%% START OF SUPPLEMENT %%%%%%%%%%%%%%%

% Figures, tables, equations and pages in the supplement are numbered S1, S2 etc.
\renewcommand{\thefigure}{S\arabic{figure}}
\renewcommand{\thetable}{S\arabic{table}}
\renewcommand{\theequation}{S\arabic{equation}}
\renewcommand{\thepage}{S\arabic{page}}
\setcounter{figure}{0}
\setcounter{table}{0}
\setcounter{equation}{0}
\setcounter{page}{1} % not 0 as \newpage already started a supplementary page
% References continue the numbering from the main text.
% Supplementary Notes are numbered as in the IOP version, so that
% `Supplementary \ref{...}' still reads `Supplementary Note N.'
\renewcommand{\thesubsubsection}{Note \arabic{subsubsection}.}
\setcounter{subsubsection}{0}

%%%%%%%%%%%%%%%% SUPPLEMENT TITLE PAGE %%%%%%%%%%%%%%%

\begin{center}
\section*{Supplementary Materials for\\ \scititle}

Long~Cheng$^{\ast}$,
Alexander~Holm~Ohrt,
Oliver~Nagy,
Yu~Zhao,
Chenyu~Wang,
Yaxian~Wang,
Sheng~Meng,
Guohui~Li,
Qinying~Pan,
Xingyu~Huang,
Sanshui~Xiao,
Sijie~Zhang,
Guoqian~Liao,
Yutong~Li,
Xuan~Luo,
Yuping~Sun,
Kaibo~Zheng,
Peter~Uhd~Jepsen,
Binbin~Zhou$^{\ast}$\\
\small$^\ast$Corresponding authors. Email: longch@dtu.dk; zhou@dtu.dk
\end{center}

\subsubsection*{This PDF file includes:}
Supplementary Text\\
Figures S1 to S8

\subsection*{Supplementary Notes}

\vspace{1em}
\begin{enumerate}
\setlength{\itemsep}{0.5em}  % control distance vertically
  \item Structural Characterization of NiPS$_3$ Single Crystals                          \dotfill \pageref{si:struct}
  \item Electronic Band Structure and Orbital Hybridization                              \dotfill \pageref{si:band}
  \item Crystal-field Splitting and Steady-State Optical Transitions                     \dotfill \pageref{si:cfs}
  \item Phonon Dispersion and Projected Density of States                                \dotfill \pageref{si:phonon}
  \item Power-density Dependence of the PL Spectrum                                      \dotfill \pageref{si:fluence}
  \item Excitation-energy Dependence and Thermally Activated Localization Dynamics       \dotfill \pageref{si:1dt}
  \item Extraction of Equilibrium and Transient THz Optical Constants                    \dotfill \pageref{si:pgfm_and_ptmm}
  \item Spectro-temporal Mapping and Supplementary Features                              \dotfill \pageref{si:spectro}
  \item Parity Considerations for Cooperative Phonon-assisted Self-trapping              \dotfill \pageref{si:srSTE}
\end{enumerate}

\newpage

%%%%%%%%%%%%%%%% SUPPLEMENTARY TEXT %%%%%%%%%%%%%%%

\subsection*{Supplementary Text}

\subsubsection{Structural Characterization of NiPS$_3$ Single Crystals}\label{si:struct}

\begin{figure} % Do NOT use \begin{figure*}
\centering
\includegraphics[width=0.8\linewidth]{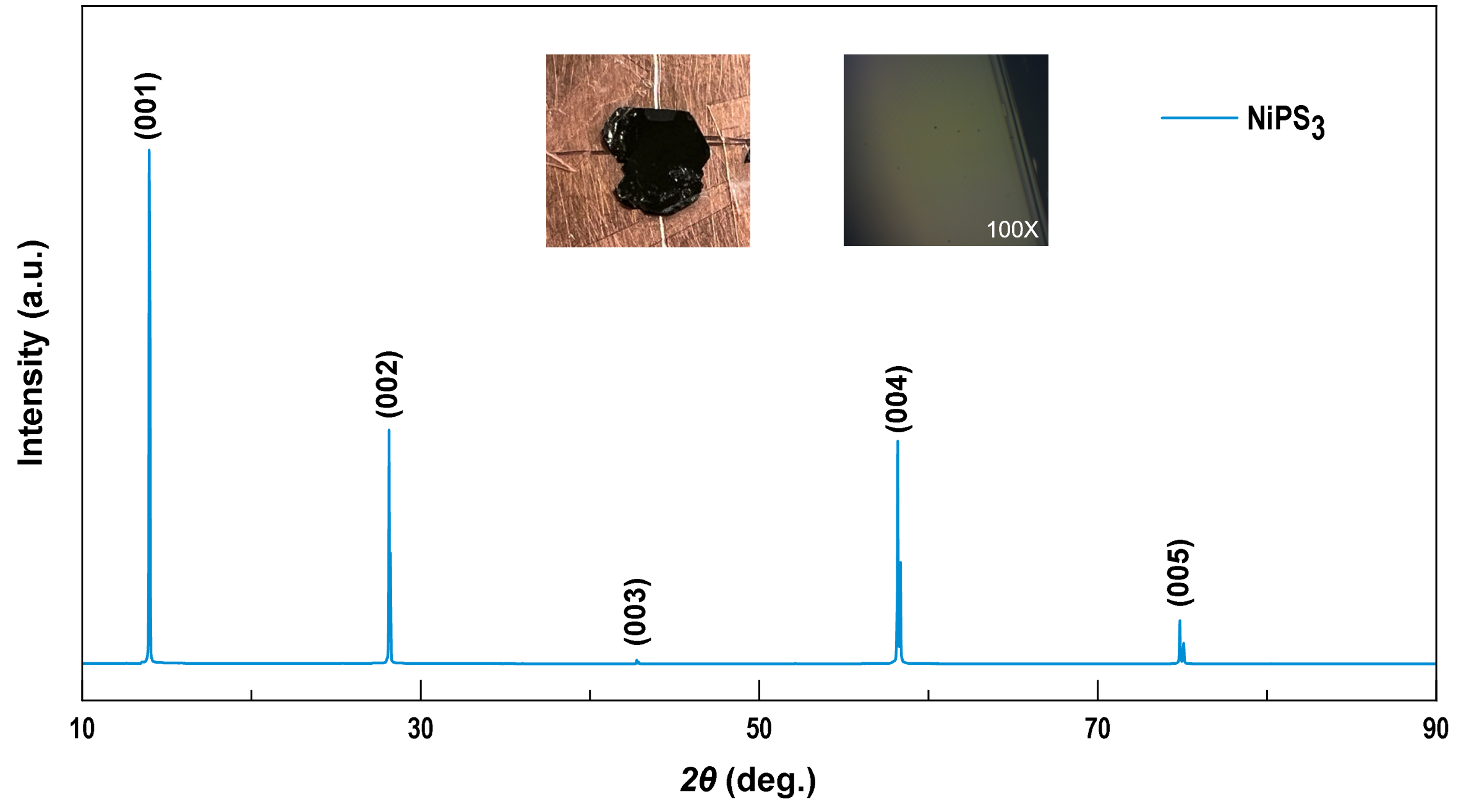}
\caption{\textbf{X-ray diffraction and surface morphology of the NiPS$_3$ single crystal.} X-ray diffraction measured from the exfoliated basal plane exhibits exclusively $(00l)$ Bragg reflections, confirming the high crystallinity and phase purity of the crystal. Insets: Photograph of a representative bulk NiPS$_3$ (left) and a 100$\times$ optical microscope image of the mechanically exfoliated basal plane (right), revealing a smooth, optically flat, and highly reflective surface.}
\label{fig:figS1.png}
\end{figure}

\textbf{Structural integrity and phase purity.} To assess the crystalline quality and phase purity of the NiPS$_3$ single crystals synthesized by chemical vapor transport (CVT), X-ray diffraction (XRD) measurements were performed on the mechanically exfoliated flakes. As shown in Fig.~\ref{fig:figS1.png}, the diffraction pattern is dominated by the $(00l)$ family of Bragg reflections, spanning the $(001)$ to $(005)$ peaks, consistent with the layered crystal structure and the preferred orientation of the basal plane. No impurity-related diffraction peaks are observed, confirming the high phase purity of the synthesized crystals. The optical microscopy image (Fig.~\ref{fig:figS1.png}, right inset) reveals a smooth and uniform surface obtained by cleavage along the van der Waals basal plane. This absence of visible surface damage or large-scale morphological defects indicates the excellent surface quality and provides an optically flat interface for ultrafast spectroscopy measurements.

\subsubsection{Electronic Band Structure and Orbital Hybridization}\label{si:band}

\textbf{Orbital-projected electronic structure and density of states} The electronic structure of NiPS$_3$ was calculated within density functional theory (DFT) using the Vienna \textit{Ab initio} Simulation Package (VASP)\cite{vasp1}. To capture the correlated electronic states and charge-transfer-insulator character of NiPS$_3$, structural optimization and electronic structure calculations were performed using the strongly constrained and appropriately normed (SCAN) meta-generalized gradient approximation (meta-GGA)\cite{metaGGA, NPS_Struc}. Previous studies have shown that the SCAN functional reproduces both the experimental band gap and the localized electronic states of this correlated van der Waals antiferromagnet \cite{abinitio}.

\begin{figure} % Do NOT use \begin{figure*}
\centering
\includegraphics[width=0.8\linewidth]{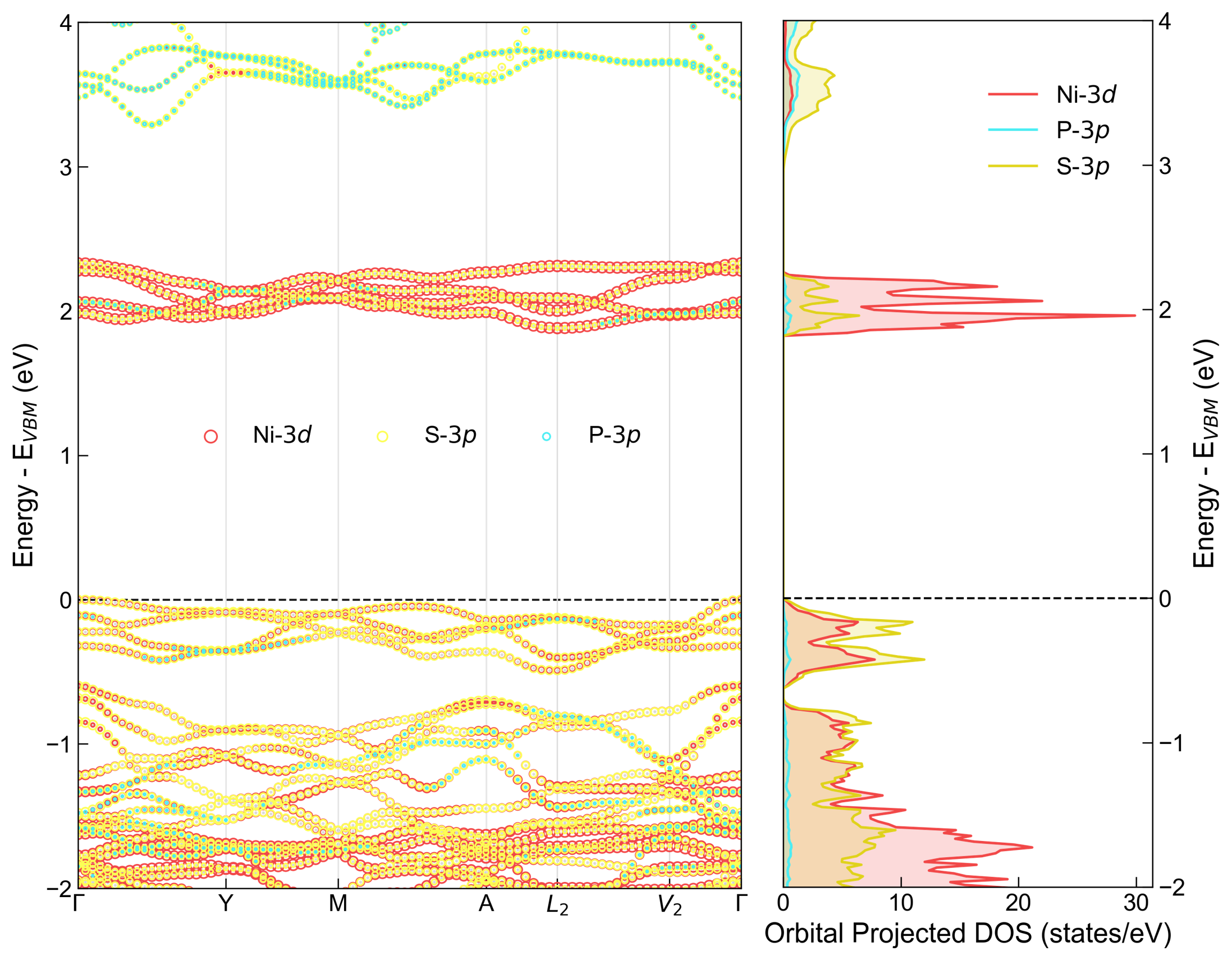}
\caption{\textbf{Electronic structure and orbital character of NiPS$_3$.} \textbf{Left:} Orbital-projected electronic band structure calculated using the strongly constrained and appropriately normed (SCAN) meta-GGA functional. The size and color of the circles indicate the fractional contributions of Ni $3d$ (red), S $3p$ (yellow), and P $3p$ (cyan) orbitals. The valence-band maximum (VBM) is aligned to 0 eV (horizontal dashed line). \textbf{Right:} Corresponding orbital-projected partial density of states (PDOS). The valence-band maximum region is composed predominantly of strongly hybridized S $3p$ and Ni $3d$ states, whereas the conduction-band minimum is dominated by localized Ni $3d$ orbitals. This orbital configuration is consistent with the charge-transfer-insulator character of NiPS$_3$ and provides the microscopic basis for the ligand-to-metal charge-transfer excitation ($\mathrm{S}\,3p \rightarrow \mathrm{Ni}\,3d$) employed in the ultrafast measurements. The calculated charge-transfer gap is approximately 1.79 eV.}
\label{fig:figS2.png}
\end{figure}

Fig. \ref{fig:figS2.png} shows the orbital-projected band structure (left panel) and corresponding partial density of states (PDOS, right). The projections reveal the characteristic charge-transfer electronic structure of NiPS$_3$. The valence band maximum (VBM, aligned to 0 eV) consists predominantly of strongly hybridized S $3p$  and Ni $3d$ states, as evidenced by the substantial overlap of the corresponding orbital weights. Deeper valence bands originate primarily from non-bonding sulfur $3p$ states. This electronic configuration is consistent with a correlated ground state containing mixed $d^8$ and $d^9\underline{L}$ character, where $\underline{L}$ denotes a ligand hole on the sulfur sublattice. In contrast, the conduction band minimum (CBM) is dominated by localized Ni $3d$ states, primarily of $e_g$ character, with comparatively weaker contributions from sulfur. This orbital configuration provides the microscopic basis for the ligand-to-metal charge-transfer (LMCT) excitation employed in the ultrafast measurements. Optical excitation at 1.91 eV promotes electrons from the hybridized S $3p$--Ni $3d$ states near the VBM into unoccupied Ni $3d$ ($e_g$) states at the CBM. Because the lobes of the $e_g$  orbitals are directed toward the surrounding sulfur ligands within the NiS$_6$ octahedra, this excitation induces substantial redistribution of charge density along the Ni--S bonds and strongly perturbs the local bonding environment. Furthermore, the localized character of the Ni $3d$ states gives rise to a relatively narrow conduction bandwidth. Within strong-coupling polaron models, the effective electron--phonon coupling strength increases as the electronic bandwidth decreases \cite{CTmaterial_e-phon_couple}. The calculated effective conduction bandwidth ($W \approx 192$~meV) inherently limits carrier itinerancy. Consequently, photoexcitation into the Ni $3d$ manifold is expected to enhance carrier--lattice interactions and promote lattice-assisted carrier localization. These considerations provide a microscopic framework for understanding the pronounced phonon renormalization and exciton localization dynamics observed in the time-resolved measurements.

\subsubsection{Crystal-field Splitting and Steady-State Optical Transitions} \label{si:cfs}

\begin{figure} % Do NOT use \begin{figure*}
\centering
\includegraphics[width=\linewidth]{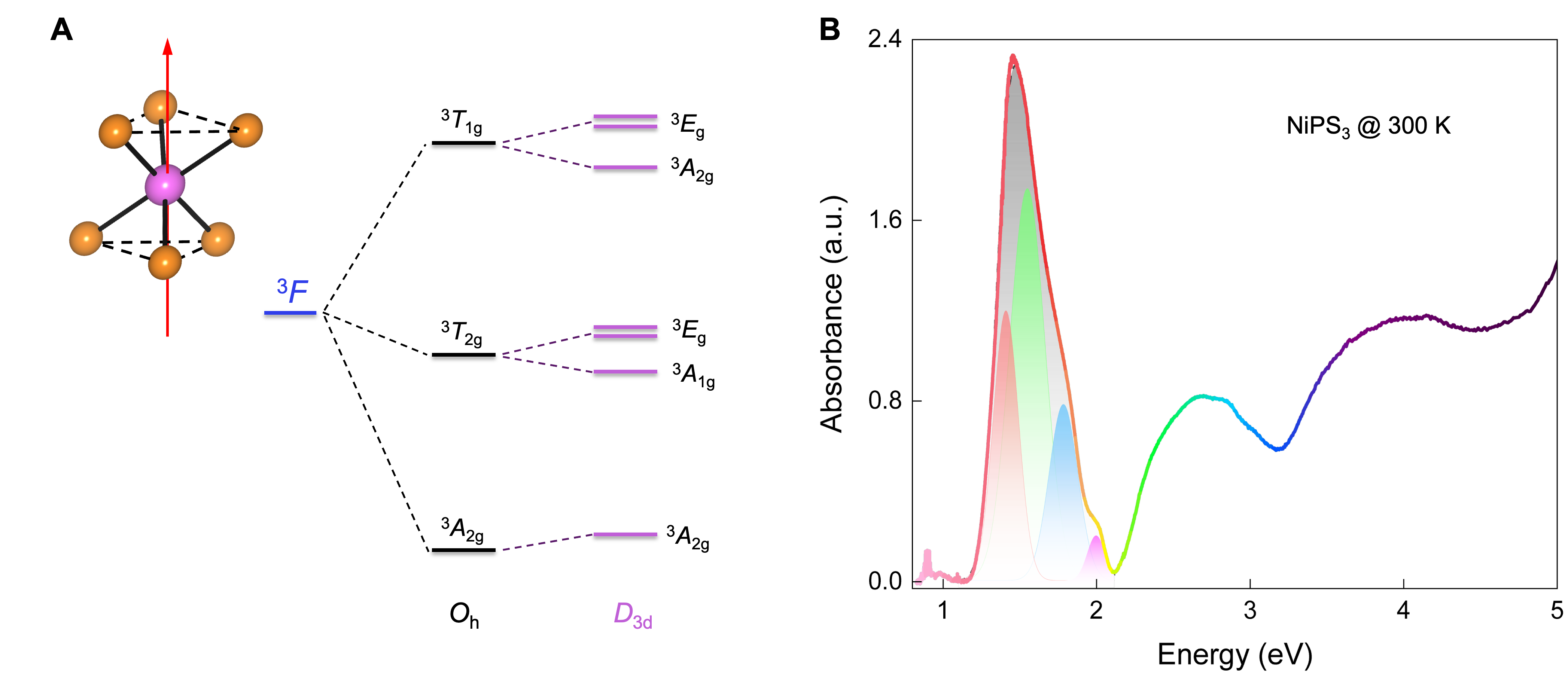}
\caption{\textbf{Crystal-field splitting and room-temperature optical absorption of NiPS$_3$.} (\textbf{A}) Schematic illustration of the local NiS$_6$ octahedral coordination and the associated crystal-field splitting of the Ni$^{2+}$ ($3d^8$) multiplet states. The intrinsic trigonal distortion lowers the local symmetry from $O_h$ to $D_{3d}$, lifting the degeneracy of the crystal-field levels and giving rise to the additional splitting observed in the optical spectra.  (\textbf{B}) Room-temperature optical absorption spectrum of NiPS$_3$ together with a multi-peak decomposition. Weak absorption features at $\sim$0.9 and $\sim$1.1 eV are assigned to the parity-forbidden on-site  $d-d$ transitions from the $^3A_{2g}$ ground state to the split $^3A_{1g}$ and $^3E_g$ levels derived from the  $^3T_{2g}$ manifold. The broad absorption continuum spanning 1.2--2.1 eV is decomposed into four components: an excitonic resonance centered near $\sim$1.4 eV (pink), a ligand-to-metal charge-transfer (LMCT) continuum near $\sim$1.6 eV (green), and higher-energy crystal-field excitations near $\sim$1.78~eV (cyan) and $\sim$2.0~eV (purple) associated with states derived from the $^3T_{1g}(F)$ manifold. Additional features near $\sim$2.6 and $\sim$4.0 eV are attributed to higher-lying crystal-field excitations and charge-transfer transitions \cite{RN3556}.
 }
\label{fig:figS3.png}
\end{figure}

\begin{figure} % Do NOT use \begin{figure*}
\centering
\includegraphics[width=\linewidth]{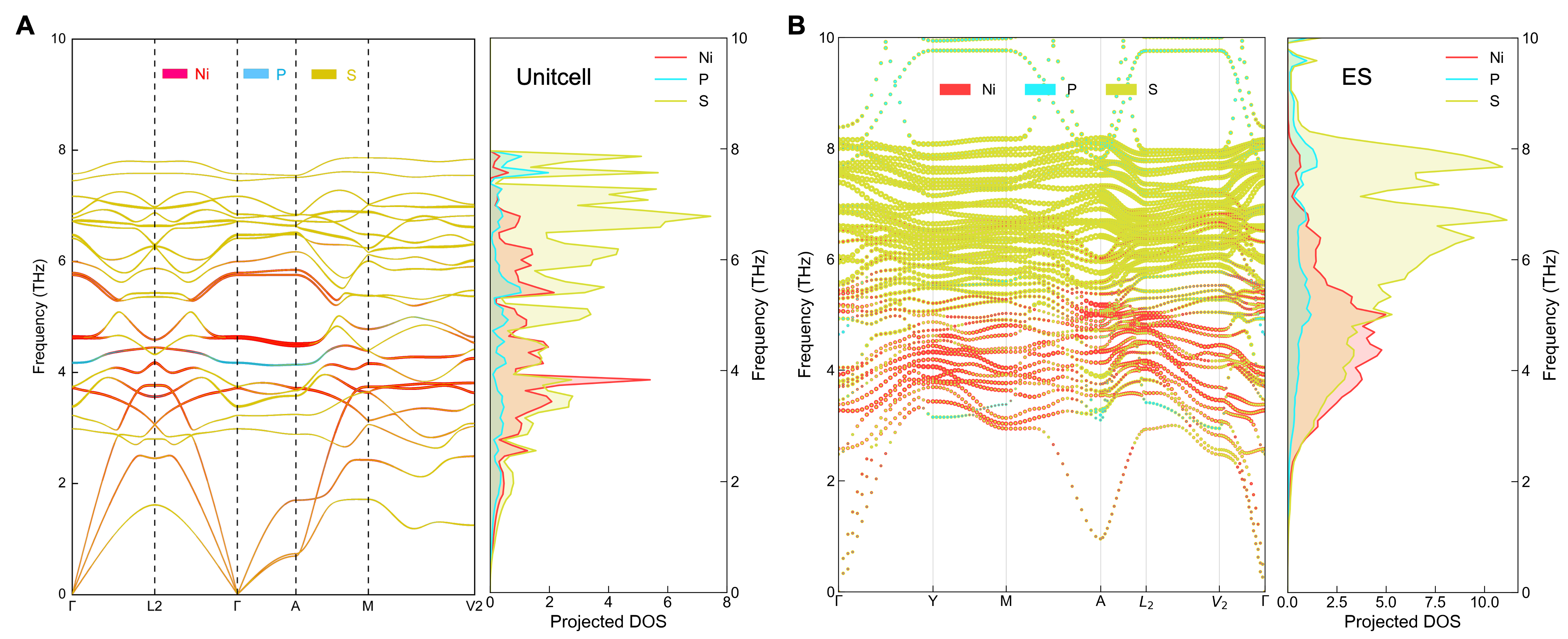}
\caption{\textbf{Phonon dispersion and projected density of states of NiPS$_3$.} (\textbf{A}) Phonon dispersion of the primitive cell calculated along the high-symmetry directions of the Brillouin zone. Marker sizes and color indicate the fractional atomic contributions from Ni (red), P (cyan), and S (yellow). The right panel shows the corresponding atom-projected phonon density of states (PDOS). (\textbf{B}) Unfolded phonon dispersion and corresponding PDOS calculated for the charge-transfer excited state. The vibrational spectrum of the photoinduced symmetry-broken state is projected back onto the equilibrium $C2/m$ Brillouin zone, enabling identification of the origin of the emergent transient modes. The calculations reveal that the weakly dispersive 2.7~THz branch, dominated by coupled Ni--S vibrations, originates from the zone-boundary phonons that become folded toward the zone center following the photoinduced structural distortion. The low group velocity of this branch gives rise to an enhanced PDOS, consistent with the transiently activated mode observed experimentally.}
\label{fig:figS4.png}
\end{figure}

\noindent\textbf{Crystal-field splitting and charge-transfer transitions} The electronic structure and optical response of NiPS$_3$ are governed by the local coordination environment of the Ni$^{2+}$ ($3d^8$) cations. As illustrated in Fig. \ref{fig:figS3.png} \textbf{A}, the intrinsic trigonal distortion lowers the local symmetry from $O_h$ to $D_{3d}$, lifting the degeneracy of the crystal-field levels. As a result, the low-lying $^3T_{2g}$ manifold splits into the $^3A_{1g}$ and $^3E_g$ states, giving rise to the weak absorption features near $\sim$0.9 and $\sim$1.1 eV, respectively. Because these intra-atomic $d-d$ transitions are parity-forbidden by the Laporte selection rule, they exhibit relatively small oscillator strengths. These weak features are resolved because the spectra were acquired in diffuse-reflectance geometry on cleaved bulk crystals, where multiple scattering lengthens the effective optical path; spectra are reported as instrument absorbance, $A=\log_{10}(1/R)$, with the specular component included. Since $R$ is bounded below by the front-surface Fresnel reflectivity once the crystal becomes opaque, strongly absorbing bands are compressed while weak sub-gap features are retained, so the weight of the crystal-field bands relative to the charge-transfer continuum in Fig. \ref{fig:figS3.png}\textbf{B} represents an upper bound. In contrast, the spectrum between 1.2 and 2.1 eV is dominated by a broad absorption continuum that can be decomposed into four overlapping components (Fig. \ref{fig:figS3.png}\textbf{B}). The feature centered near $\sim$1.4 eV is assigned to the Zhang--Rice-like excitonic resonance, whereas the broad continuum near $\sim$1.6 eV originates from the ligand-to-metal charge-transfer (LMCT) transition ($S~3p \rightarrow Ni~3d~e_g$). The onset of this charge-transfer excitation shifts from approximately 1.8 eV at low temperatures down to $\sim$1.6 eV at room temperature, reflecting the influence of lattice expansion and electron--phonon coupling on the electronic structure \cite{bandgap_300K}.

Two additional absorption features appear near $\sim$1.78 eV and $\sim$2.0 eV and are assigned to crystal-field excitations derived from the split $^3T_{1g}(F)$ manifold. Compared with the lower-energy $^3T_{2g}$ transitions, these excitations exhibit substantially larger oscillator strengths and broader linewidths. Such enhancement is commonly attributed to intensity borrowing from the nearby optically allowed charge-transfer continuum \cite{high_orderdd,enhance_dd_absorp_CT}. Because these states lie within or close to the LMCT absorption band, hybridization between sulfur $3p$ and nickel $3d$ orbitals partially relaxes the dipole-selection constraints and enhances their optical activity \cite{exciton}. The corresponding linewidth broadening is consistent with stronger coupling to the surrounding electronic and lattice continua. This optical landscape is particularly relevant for the ultrafast measurements presented in the main text. Excitation at 1.91 eV directly accesses the LMCT continuum (Fig.~\ref{fig:figS3.png}\textbf{B}), generating high-energy charge-transfer carriers within a strongly hybridized electronic manifold. Although the split $^3T_{1g}(F)$ states lie nearby, their oscillator strengths arise from mixing with the allowed charge-transfer continuum rather than from the intra-atomic $d^8$ configuration. Moreover, a purely intra-atomic $d-d$ excitation redistributes charge within a single Ni site and is charge-neutral and cannot generate the mobile-carrier response that dominates the transient terahertz signal at early delays. Any residual crystal-field absorption at 1.91 eV would therefore open a parallel, non-conductive relaxation channel contributing neither to the measured conductivity nor to the subsequent localization. The associated redistribution of charge density along the Ni--S bonds provides favorable conditions for strong carrier--lattice coupling and the subsequent localization dynamics observed in the time-resolved experiments.

\subsubsection{Phonon Dispersion and Projected Density of States}\label{si:phonon}

\noindent\textbf{Phonon dispersion and density of states.} Phonon dispersion relations and the corresponding density of states (DOS) were calculated using the finite-displacement method as implemented in the VASP and Phonopy packages\cite{phonopy}. The interatomic force constants were obtained from a $2 \times 2 \times 2$ supercell. Electronic exchange and correlation were treated using the Perdew--Burke--Ernzerhof (PBE) functional with a Hubbard correction of $U=4.0$ eV applied to the Ni $3d$ states in order to reproduce the localized electronic character and zigzag antiferromagnetic ground state of NiPS$_3$.

The calculated atom-projected phonon dispersion identifies the microscopic origin of the lattice modes observed experimentally. The branch at $\sim$5.7 THz corresponds to the infrared-active $B_u$ mode and is dominated by vibrations of the Ni atoms within the NiS$_6$ octahedra \cite{RN3592}. In contrast, the optical phonon branches near $\sim$8.0 THz are primarily associated with sulfur displacements. This frequency range contains the infrared-active $A_u$ mode arising from Davydov splitting of the corresponding $A_g$ vibration\cite{ir_active}. The mode involves collective sulfur motions that modulate the Ni--S octahedral environment and dynamically perturb the crystal-field splitting of the Ni $3d$ orbitals\cite{ir_active,ir_active1,elecphonon}. Such lattice distortions are expected to couple efficiently to charge-transfer excitations because they directly modify the local Ni--S bonding geometry.

Beyond the zone-center infrared-active modes, the calculations reveal a weakly dispersive phonon branch near $\sim$2.7 THz along several zone-boundary directions, including the $Y$--$M$ and $L_2$--$V_2$ paths. The small group velocity of these modes gives rise to an enhanced phonon density of states near this frequency, characteristic of a flat phonon branch. The atom-projected DOS indicates that these vibrations involve coupled Ni--S displacements. In the C-centered NiPS$_3$ lattice, the zone-boundary phonons at the Y point are antisymmetric under the centering translation, so the dipoles of translation-related atoms cancel and leave these modes dark in equilibrium. A localized exciton could remove this translation at its site, making the two centering-related halves of the cell inequivalent and rendering these modes optically active. Within the $P2_1/m$ subgroup that retains inversion through the midpoint of a nearest-neighbor Ni--Ni bond, the 2.7 THz mode transforms as $A_u$. The dipole of each mode scales with the relative displacement between these two halves in its equilibrium eigenvector. By this measure, the 2.7 THz mode ranked first among the 30 Y-point modes. Due to the centering translation belonging to the crystal structure rather than the magnetic order, the same activation operates at 300 K, above the Néel temperature.

To explicitly investigate the origin of the transiently activated modes, we calculated the structural and vibrational response of the charge-transfer excited state. Promoting an electron from the hybridized valence-band maximum to the localized Ni $3d$ conduction-band minimum induces a localized structural relaxation. Rather than a macroscopic phase transition, this local distortion microscopically mimics a primitive-cell doubling within the spatial extent of the exciton. By folding the phonon spectrum of this extended cell onto the equilibrium $C2/m$ Brillouin zone (Fig.~\ref{fig:figS4.png}\textbf{B}), the origin of the emergent vibrational branches could be identified through their spectral weights. The calculations reveal that the transiently emergent modes originate from weakly dispersive zone-boundary phonons and acquire transient optical activity. For instance, the emergent 4.5 THz mode is attributed to a second zone-boundary branch, in which Ni displacements couple to the (P$_2$S$_6$)$^{4-}$ bipyramidal units. This branch is activated by the same loss of the C-centering translation, but less efficiently than the 2.7 THz case. Notably, the 2.7~THz branch exhibits a substantially stronger transient dielectric response than would be expected solely from its phonon density of states. This observation suggests that the nonequilibrium terahertz response is governed not only by the available vibrational phase space, but also by the strength of the coupling between specific lattice coordinates and the transient charge-transfer dipole. The strong Ni--S participation of the 2.7~THz vibration therefore also makes it a particularly sensitive indicator of the lattice response during exciton localization.

\subsubsection{Power-density Dependence of the PL Spectrum}\label{si:fluence}

\begin{figure} % Do NOT use \begin{figure*}
\centering
\includegraphics[width=\linewidth]{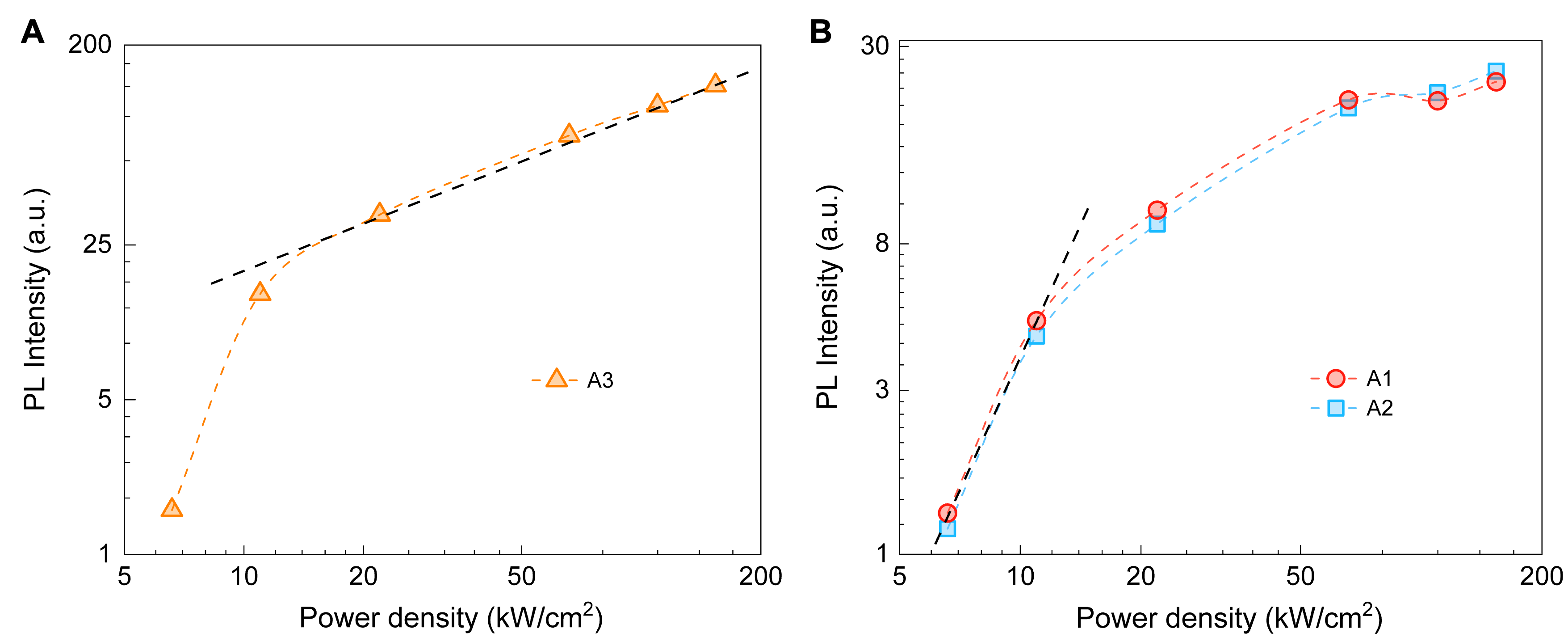}
\caption{\textbf{Power density dependence of the photoluminescence emission in NiPS$_3$.} 
(\textbf{A}) Log--log plot of the integrated photoluminescence (PL) intensity of the 1.295 eV emission band as a function of excitation power density. 
(\textbf{B}) Corresponding log--log plots of the integrated intensities of the 1.409 eV (A$_1$) and 1.363 eV (A$_2$) excitonic emission features. Above an excitation threshold of approximately 11 kW/cm$^{2}$, both excitonic peaks exhibit pronounced sublinear behavior, consistent with the onset of exciton--exciton interactions and saturation effects. Concurrently, the 1.295 eV emission band emerges and follows an approximately $P^{1/2}$ power-law dependence over the measured excitation range. Black dashed lines indicate fits to the corresponding power-law scaling.}
\label{fig:figS5.png}
\end{figure}

\noindent\textbf{High-density exciton dynamics and exciton--exciton annihilation.} To further investigate the population dynamics at elevated excitation densities, we measured the excitation-power dependence of the steady-state photoluminescence (PL) spectra. As shown in Fig.~\ref{fig:figS5.png}, the emission exhibits a crossover near an excitation power density of $\sim$11 kW/cm$^{2}$. Below this threshold, the integrated intensities of the 1.409 eV (A$_1$) and 1.363 eV (A$_2$) emission features scale approximately linearly with excitation power, consistent with a regime dominated by independent excitonic recombination. In contrast, the low-energy emission band near 1.295 eV is negligible in this regime.

Above this threshold, both excitonic emission features exhibit pronounced sublinear behavior, indicating the onset of density-dependent exciton interactions. Such behavior is consistent with exciton--exciton annihilation (EEA), which becomes increasingly important as the average exciton separation decreases. Within a simple rate-equation framework, the exciton population density $n$ obeys 
\begin{equation}
\frac{dn}{dt}=\alpha P-\gamma n^2,
\label{eq1}
\end{equation}

\noindent where $\alpha$ is the excitation efficiency, $P$ is the pump power density, and $\gamma$ is the EEA coefficient. Under steady-state conditions ($dn/dt=0$), the exciton density scales as $n\propto P^{1/2}$.

Concurrently with the onset of EEA, the 1.295 eV emission band becomes clearly visible and exhibits an approximately $P^{1/2}$ power-law dependence over the measured excitation range up to 154 kW/cm$^{2}$. The correspondence between the observed scaling and the EEA-limited exciton population suggests that this low-energy emission is populated through density-dependent relaxation channels that become active at high excitation densities. While the microscopic origin of this feature cannot be uniquely determined from the present measurements, its emergence only in the high-density regime is consistent with a localized excited state that is efficiently fed by exciton--exciton interactions and subsequent energy relaxation processes.

\subsubsection{Excitation-energy Dependence and Thermally Activated Localization Dynamics}\label{si:1dt}

\noindent Optical excitation at 800 nm (1.55 eV) predominantly addresses the Zhang--Rice excitonic resonance and deposits relatively little excess energy into the lattice. Under these conditions, the transient response remains largely positive and carrier-like throughout the measured time window. In contrast, excitation at 650 nm (1.91 eV) directly accesses the ligand-to-metal charge-transfer (LMCT) continuum, generating initially unbound carriers and depositing substantially greater excess energy into the coupled charge--lattice system. The resulting dynamics evolve toward a localized dielectric state, highlighting the strong dependence of the nonequilibrium response on the excitation pathway.

\begin{figure} % Do NOT use \begin{figure*}
\centering
\includegraphics[width=\linewidth]{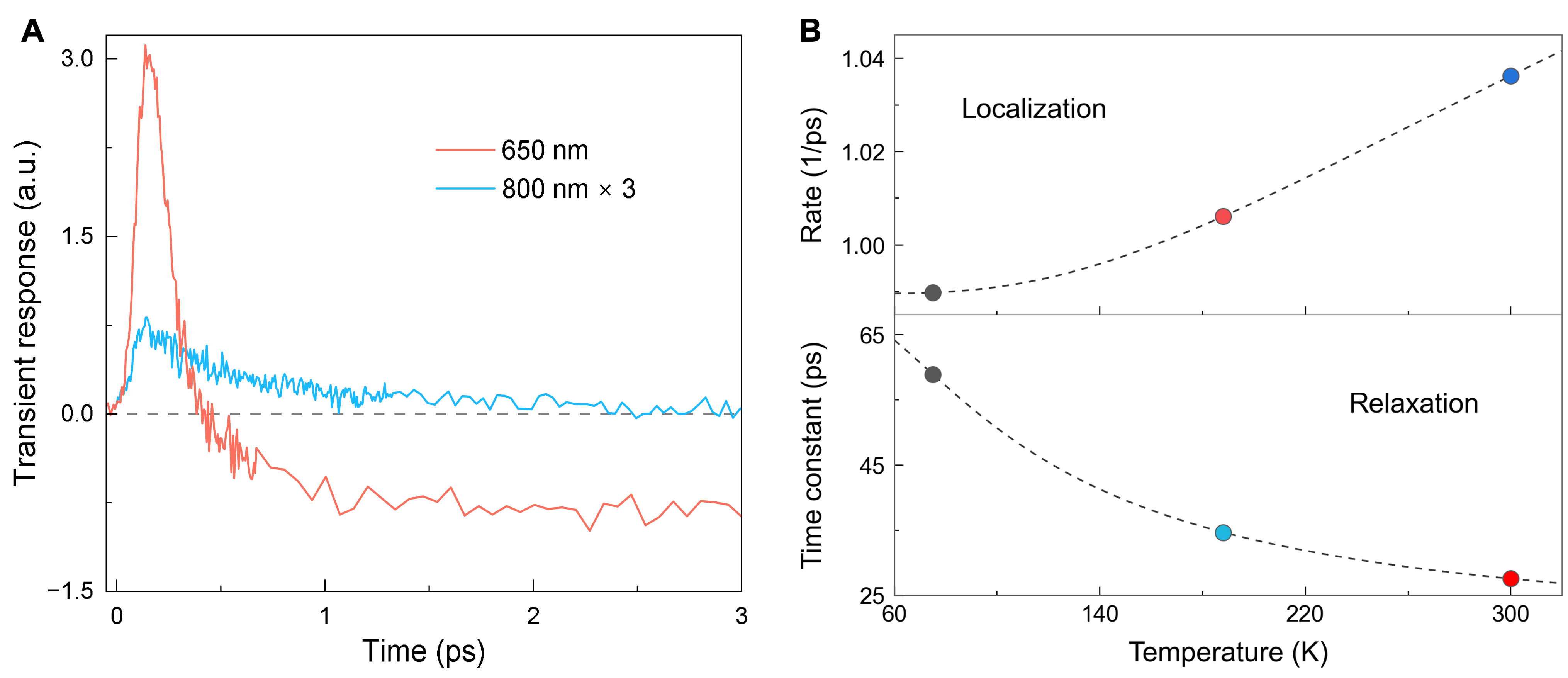}
\caption{\textbf{Excitation-energy dependence and temperature dependence of the localization dynamics.} (\textbf{A}) Time-resolved terahertz reflectivity dynamics measured at 300 K following excitation at 650 nm (1.91 eV, red) and 800 nm (1.55 eV, blue). The 800 nm trace is multiplied by a factor of three for clarity. Both measurements were performed at the same incident pump fluence of 1.3 mJ$\cdot$cm$^{-2}$. (\textbf{B}) Temperature dependence of the localization rate (upper panel) and the relaxation time constant extracted from multi-exponential fits to the OPTP dynamics (lower panel) at 75, 188, and 300 K. Dashed lines represent Arrhenius analysis employed to extract the localization barrier $E_b$ (upper panel) and effective activation energy $E_{\mathrm{eff}}$ (lower panel) of the subsequent relaxation process.}
\label{fig:figS6.png}
\end{figure}

The temperature dependence of the transient dynamics provides further insight into the localization process. Both the formation rate of the negative transient response ($k_{\mathrm{form}}$) and the amplitude of the localized component ($|A|$) are well described by Arrhenius-type expressions,
$k_{\mathrm{form}}(T)=k_0+\nu\exp(-E_b/k_\mathrm{B}T)$ and $|A(T)|=A_{\mathrm{res}}+A_{\mathrm{th}}\exp(-E_A/k_\mathrm{B}T)$, where $k_{\mathrm{form}}$ is the build-up rate of the negative response following exciton formation, with $k_0$ and $A_{\mathrm{res}}$ denoting temperature-independent offsets\cite{STEbook}. Fitting $k_{\mathrm{form}}$ (upper panel of Fig.~\ref{fig:figS6.png}\textbf{B}) and $|A|$ (Fig.~\ref{fig:fig2.png}\textbf{E}, inset) yields $E_b\approx45\pm2$ meV and $E_A\approx43\pm2.3$ meV, respectively. The two energies are numerically close but physically distinct, with $E_b$ setting the activation barrier that governs the localization rate and $E_A$ describing the thermally activated growth of the localized population.

The subsequent recovery of the negative transient response reflects the depletion of the localized excitonic population. Unlike the initial localization process, this slower relaxation involves multiple competing channels, including thermal detrapping, non-radiative recombination, and transfer to intermediate dark states. Consequently, the activation energy extracted from the recovery dynamics ($E_{\mathrm{eff}}\approx24.5$ meV) should be regarded as an effective phenomenological energy scale rather than a unique microscopic detrapping barrier. This energy characterizes the thermal destabilization of the localized state and provides a measure of its stability against thermally activated relaxation processes. Given that $E_b$ is comparable to $k_BT$ at room temperature, localization does not exhaust the free exciton population. Both configurations therefore remain populated and retain their own radiative channel, which accounts for the two excitonic emission features resolved in the steady-state PL spectrum (Fig. \ref{fig:fig1.png}\textbf{D}). Their relative weights are consistent with the fraction of the photoexcited population transferred into the long-lived localized state (Fig. \ref{fig:fig2.png}\textbf{B}). Such coexistence is well documented in molecular crystals \cite{STEbook}.

\subsubsection{Extraction of Equilibrium and Transient THz Optical Constants}\label{si:pgfm_and_ptmm}

\textbf{Extraction of intrinsic terahertz optical constants via parameterized global forward modeling (PGFM).} The extraction of complex optical constants from terahertz time-domain spectroscopy (THz-TDS) is commonly performed through direct numerical inversion of the measured transfer function. For thin van der Waals crystals, however, multiple internal reflections generate overlapping Fabry--Pérot echoes that complicate this procedure. Conventional approaches often require temporal windowing of the time-domain waveform, which can reduce low-frequency spectral fidelity, or become susceptible to phase-unwrapping ambiguities near strong phonon resonances. To overcome these limitations, we employ the parameterized global forward modeling (PGFM) approach \cite{pgfm}. Rather than extracting the optical constants independently at each frequency, PGFM describes the dielectric response using a physically constrained model and determines the corresponding parameters through global optimization. The measured complex transfer function is reproduced by explicitly accounting for the coherent Fabry--Pérot reflections within the sample, thereby enabling a self-consistent determination of the equilibrium dielectric response across the full spectral range. For NiPS$_3$, the equilibrium dielectric function is represented by a hybrid Drude--Smith and double-Lorentz-oscillator model:

\begin{equation}
\tilde{\varepsilon}(\omega)=\varepsilon_\infty+\sum_{k=1}^{2}\frac{A_k}{\Omega_k^2-\omega^2-i\gamma_k\omega}
+\frac{i\tilde{\sigma}_{DS}(\omega)}{\varepsilon_0\omega},
\end{equation}
\vspace{0.5em}

\noindent where $\varepsilon_\infty$ denotes the high-frequency dielectric constant. The Lorentz oscillators describe the infrared-active phonons, with $A_k$, $\Omega_k$, and $\gamma_k$ corresponding to the oscillator strength, resonance frequency, and linewidth, respectively. The Drude--Smith contribution accounts for any residual carrier response in the equilibrium state.

\begin{figure} % Do NOT use \begin{figure*}
\centering
\includegraphics[width=\linewidth]{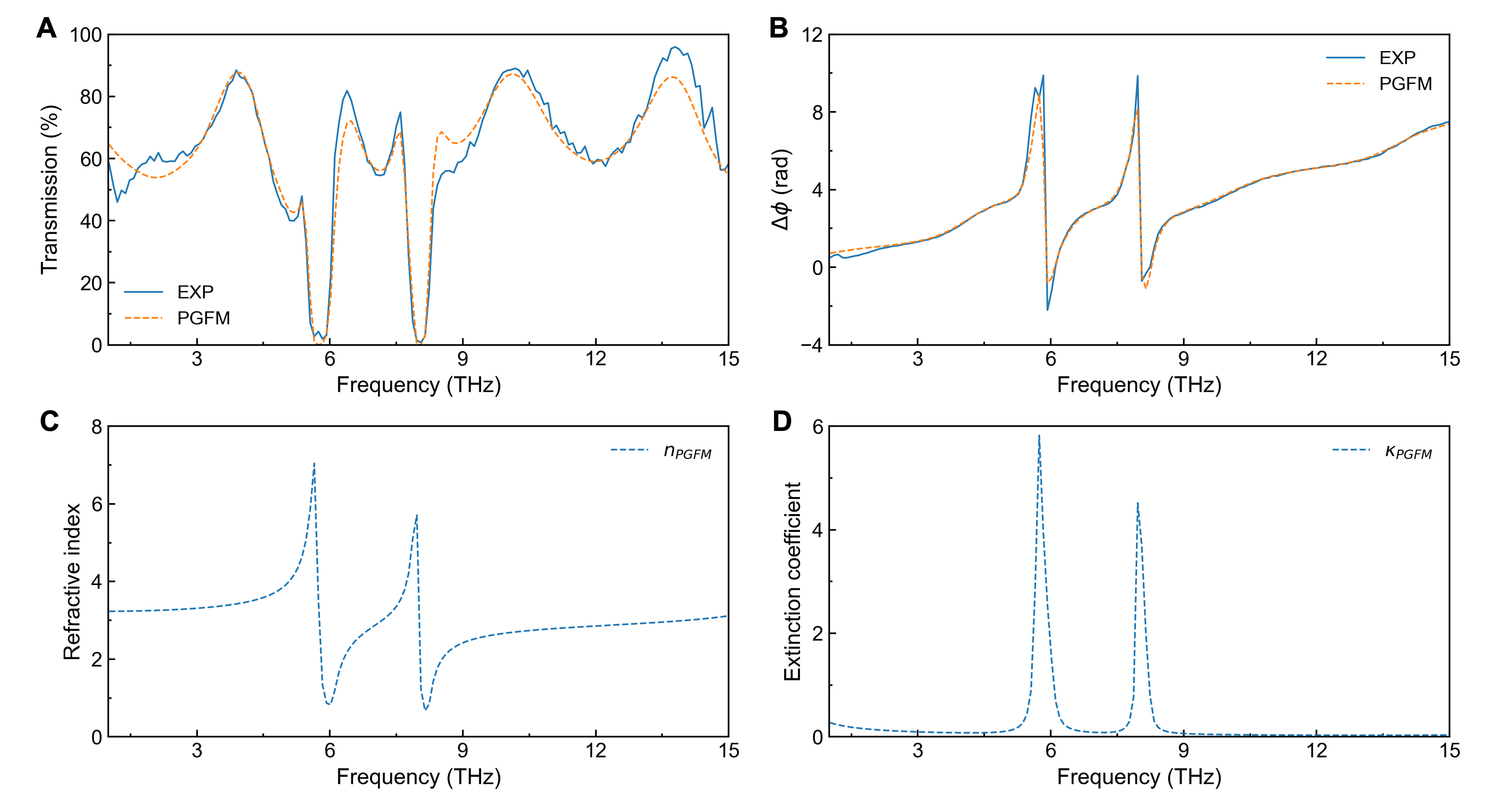}
\caption{\textbf{Extraction of equilibrium terahertz optical constants of NiPS$_3$ using parameterized global forward modeling (PGFM)\cite{pgfm}.} (\textbf{A}) Transmittance and (\textbf{B}) phase-difference ($\Delta\phi$) spectra measured by steady-state THz time-domain spectroscopy at room temperature. Solid blue curves show the experimental data, including Fabry--Pérot interference arising from multiple internal reflections within the crystal. Dashed orange curves represent the corresponding PGFM fits obtained using a hybrid Drude--Smith and double-Lorentz-oscillator dielectric model. (\textbf{C}) Refractive index ($n$) and (\textbf{D}) extinction coefficient ($\kappa$) extracted from the optimized dielectric function. By explicitly incorporating Fabry--Pérot reflections and simultaneously fitting the amplitude and phase response, PGFM provides a self-consistent description of the optical properties across the full measured spectral range, including the strong phonon resonances near 5.7 and 8.0 THz. The resulting equilibrium optical constants serve as the reference parameters for the perturbed transfer-matrix method (PTMM) analysis of the photoinduced terahertz response.}
\label{fig:figS7.png}
\end{figure}

The complex refractive index is obtained from $\tilde{n}(\omega)=\sqrt{\tilde{\varepsilon}(\omega)}$. The corresponding complex transmission function is then constructed by explicitly including multiple internal reflections within a crystal of thickness $d$,

\begin{equation}
\begin{aligned}
\tilde{H}_{\mathrm{sim}}(\omega) 
= \tilde{t}_{12}\tilde{t}_{23}\exp\left(i\frac{\omega\tilde{n}d}{c}\right)
  \sum_{m=0}^{\infty}\left[\tilde{r}_{21}\tilde{r}_{23}\exp\left(i\frac{2\omega\tilde{n}d}{c}\right)\right]^{m}
= \frac{\tilde{t}_{12}\tilde{t}_{23}\exp\left(i\frac{\omega\tilde{n}d}{c}\right)}
       {1-\tilde{r}_{21}\tilde{r}_{23}\exp\left(i\frac{2\omega\tilde{n}d}{c}\right)},
\end{aligned}
\label{eq:H_sim}
\end{equation}

\noindent where $\tilde{t}_{ij}$ and $\tilde{r}_{ij}$ are the Fresnel transmission and reflection coefficients at the corresponding interfaces, and $c$ is the speed of light. This expression is equivalent to summing all coherent Fabry--Pérot contributions to infinite order. The model parameters are optimized by minimizing the difference between the simulated transfer function $\tilde{H}_{\mathrm{sim}}(\omega)$ and the experimentally measured transfer function $\tilde{H}_{\mathrm{exp}}(\omega)$. To improve the robustness of the extraction, the fitting procedure was cross-validated using NiPS$_3$ crystals of different thicknesses. The resulting optical constants reproduce both the amplitude and phase spectra across the full measured frequency range (Fig.~\ref{fig:figS7.png}), yielding a self-consistent set of equilibrium refractive-index ($n$) and extinction-coefficient ($\kappa$) spectra. These parameters serve as the reference optical constants for the perturbed transfer-matrix method (PTMM) analysis of the photoinduced terahertz response presented in the main text.

\vspace{2em}
\noindent \textbf{Extraction of the transient dielectric response with PTMM.} Although the photoexcited layer depth of the 1.91 eV charge-transfer pump ($\delta_{pump} = \alpha^{-1} \approx 122$~nm)\cite{RN3556} is much smaller than the terahertz probe wavelength and therefore formally satisfies the thin-film limit, the extraction of transient optical parameters remains complicated by the strong frequency dependence of the equilibrium refractive index near the infrared-active phonon resonances. Under these conditions, changes in the reflected THz field contain contributions from both the intrinsic photoinduced dielectric response and the electrodynamic propagation of the probe field within the excited region. To account for these effects, we employ the perturbed transfer-matrix method \cite{PTMM}, which explicitly incorporates the depth-dependent excitation profile together with the measured equilibrium optical constants. This approach provides a self-consistent description of the interaction between the THz probe and the photoexcited layer, enabling reliable extraction of the transient dielectric function across the full measured spectral range. Within the PTMM formalism, the measured relative change in the reflected THz field, $\Delta E(\omega)/E_{ref}(\omega)$, is related to the complex photoconductivity $\Delta\tilde{\sigma}(\omega)$ through:

\begin{equation} 
\frac{\Delta\tilde{E}(\omega)}{\tilde{E}_{ref}(\omega)} = \frac{2Z_0}{\tilde{n}^2(\omega) -1} \int_{0}^{\infty} \Delta\tilde{\sigma}_{surf}(\omega) e^{-\alpha_{pump} z} e^{i \frac{2\omega\tilde{n}(\omega)}{c} z} dz \end{equation} 
\vspace{0.5em}

\noindent where $\Delta\tilde{\sigma}_{surf}(\omega)$ is the complex photoconductivity at the crystal surface, $\tilde{n}(\omega)$ is the equilibrium complex refractive index, $Z_0$ is the impedance of free space, and $c$ is the speed of light. The exponential terms account for the depth-dependent carrier distribution generated by optical absorption and the phase accumulation of the terahertz field within the photoexcited region. Evaluation of the integral yields the transient complex photoconductivity, from which the pump-induced change in the dielectric function is obtained through: 

\begin{equation} \Delta\tilde{\varepsilon}(\omega) = \Delta\varepsilon_{re}(\omega) + i\Delta\varepsilon_{im}(\omega) = i\frac{\Delta\tilde{\sigma}(\omega)}{\varepsilon_0 \omega} \end{equation} 

\noindent where $\varepsilon_0$ is the vacuum permittivity. The resulting transient dielectric spectra provide direct access to the spectro-temporal evolution of the electronic response and the phonon resonances discussed in the main text, enabling analysis of the coupling between photoexcited carriers and lattice vibrations.

\subsubsection{Spectro-temporal Mapping and Supplementary Features}\label{si:spectro}

\begin{figure} % Do NOT use \begin{figure*}
\centering
\includegraphics[width=\linewidth]{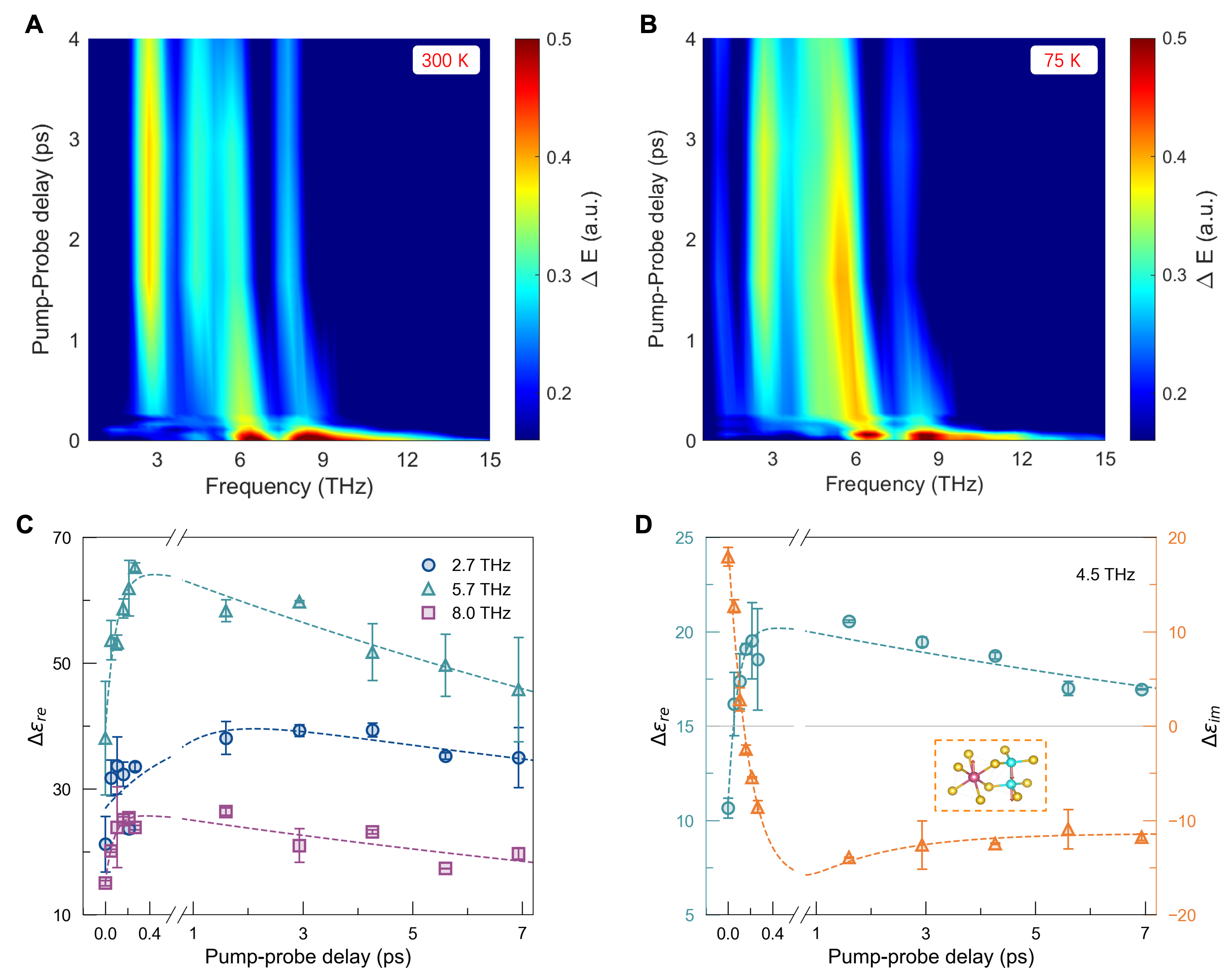}
\caption{\textbf{Spectro-temporal mapping and supplementary phonon dynamics.} 
(\textbf{A}~and~\textbf{B}) Two-dimensional maps of the pump-induced change in the reflected terahertz electric field ($\Delta E$) measured at 300 K and 75 K following 1.91 eV charge-transfer excitation. The low-temperature data reveal an additional long-lived low-frequency oscillatory feature near $\sim$1.2~THz, consistent with the antiferromagnetic magnon response reported previously.
(\textbf{C}) Temporal evolution of the transient real dielectric response ($\Delta\varepsilon_{\mathrm{re}}$) at 2.7 THz (blue circles), 5.7 THz (green triangles) and 8.0 THz (red squares). Error bars represent standard deviations and dashed curves are exponential fits. 
(\textbf{D}) Real and imaginary components of the emergent 4.5 THz mode. Insets illustrate the calculated atomic displacement patterns of the vibrational eigenmodes. Error bars represent standard deviations and dashed lines are exponential fits.
} 
\label{fig:figS8.png}
\end{figure}

\noindent \textbf{Spectro-temporal mapping and supplementary phonon dynamics.} Figure \ref{fig:figS8.png}\textbf{A, B} presents the two-dimensional spectro-temporal maps of the pump-induced changes in the reflected terahertz electric field ($\Delta E$) measured at 300 K and 75 K following 1.91 eV charge-transfer excitation. By applying PTMM to the experimental $\Delta E$ (Supplementary~\ref{si:pgfm_and_ptmm}) and subtracting the divergent low-frequency tail, we can extract the intrinsic, phonon-resolved transient dielectric response shown in Fig. \ref{fig:fig3.png}\textbf{A, B}. Because 75 K lies well below the Néel temperature of NiPS$_3$ ($T_N\approx155$ K), these measurements probe the nonequilibrium electrodynamic response within the zigzag antiferromagnetic phase. Despite the distinct magnetic environments, the overall spectro-temporal evolution remains qualitatively similar at both temperatures, indicating that the primary localization dynamics are robust against long-range magnetic ordering. In addition, the 75 K data reveal a long-lived, low-frequency oscillatory feature centered at $\sim$1.2 THz, consistent with the previously reported antiferromagnetic magnon persisting up to $T_N$\cite{RN3556}. 

The transient evolution of the intrinsic optical phonons (Fig. \ref{fig:fig3.png}) provides further insight into the lattice response following charge-transfer excitation. Following the peak itinerant carrier response at $t=0$, photoexcited carriers condense into precursor bound excitons on a timescale of $\sim$109~fs, driving the bound-state polarization recovery observed in $\Delta\varepsilon_{\mathrm{re}}$. The blueshift of the 5.7 THz ($B_u$) and 8.0 THz ($A_u$) modes in the imaginary dielectric response ($\Delta\varepsilon_{\mathrm{im}}$) is triggered after this condensation window, occurring as the excitonic wavepacket undergoes spatial collapse. This dynamic stiffening reflects the non-adiabatic electrostrictive stress exerted on local Ni--S bonds during spatial wavepacket confinement. Throughout the remaining $\sim$3 ps trajectory, the blueshifted frequencies persist without pronounced lattice softening, signaling a stabilized structural deformation that accommodates the localized state.

To microscopically isolate the roles of individual lattice coordinates in this polaronic collapse, we compare their mode-specific dielectric dynamics (Fig. \ref{fig:fig3.png}\textbf{C} and Fig. \ref{fig:figS8.png}\textbf{C,D}):
\begin{itemize}    
    \item \textbf{2.7~THz mode ($A_u$):} Coincident with the peak itinerant carrier response ($t=0$) and prior to mode activation, the imaginary dielectric component ($\Delta\varepsilon_{\mathrm{im}}$) exhibits a positive amplitude, reflecting conventional Drude-like dissipation. It crosses to negative values once the $\sim$109~fs condensation window ends and deepens as the mode is activated during the $\sim$3~ps localization. Concurrently, the real component ($\Delta\varepsilon_{\mathrm{re}}$) climbs to a peak at $\sim$3~ps (a $\sim$90\% increase relative to its itinerant-carrier peak). This electrodynamic crossover from itinerant absorption to a negative localized dielectric state marks the activation of the 2.7~THz mode as a spectroscopic indicator of exciton localization.   
    
    \item \textbf{4.5~THz mode ($B_u$):} Although sharing the same folded zone-boundary origin and a similar zero-crossing transition with the 2.7~THz mode, this coordinate exhibits a weaker dielectric response. Its imaginary component ($\Delta\varepsilon_{\mathrm{im}}$) similarly transitions from positive Drude dissipation to a negative localized state over an identical timescale of $\sim$109~fs, yet its negative excursion remains shallow. Concurrently, its real dielectric modulation ($\Delta\varepsilon_{\mathrm{re}}$) attains less than half the amplitude of the 2.7~THz mode. Lattice-dynamics calculations attribute this weaker response to the smaller dynamic dipole that this branch acquires when the C-centering translation is lost (Supplementary~\ref{si:phonon}).
    
    \item \textbf{5.7~THz mode ($B_u$):} Driven impulsively by the photoinduced dipolar field across the charge-transfer gap, this coordinate exhibits a steep real dielectric jump ($\Delta\varepsilon_{\mathrm{re}} \approx +27$), reaching its peak within the precursor exciton condensation window. This large response without progressive accumulation indicates prompt charge-transfer polarization but limited participation in the slower excitonic localization process. Moreover, its imaginary dielectric component ($\Delta\varepsilon_{\mathrm{im}}$) remains strictly positive without sign inversion, displaying a sub-100~fs transient drop and rapid upward turnaround (a kinetic bounce) before evolving into monotonic decay. This sustained positive response and kinetic bounce are consistent with continued energy exchange with the photoexcited manifold. The mode-selective kinetic bounce may originate from partial non-adiabatic reflection of the coherent lattice wavepacket involving an $A_u\otimes B_u=B_g$ cross-coupling component. Accordingly, the 5.7~THz vibration plays a non-dominant role during spatial wavepacket confinement.
    
    \item \textbf{8.0~THz mode ($A_u$):} Its real dielectric component ($\Delta\varepsilon_{\mathrm{re}}$) rises promptly within $\sim$109~fs but reaches only $\sim$33$\%$ of the 5.7 THz response. By contrast, $\Delta\varepsilon_{\mathrm{im}}$ bleaches by $\sim$70$\%$ over the same window and remains at a reduced plateau without recovery. The component-resolved dynamics distinguish this behavior from the 4.5 THz mode, whose real modulation and negative imaginary excursion are both weak. Their divergence shows that the 8.0 THz oscillator is strongly perturbed despite its limited real modulation, supporting pinning within locally distorted $\mathrm{NiS}_6$ octahedra. The sustained bleaching tracks a redistribution of oscillator spectral weight associated with changes in the local Ni–S restoring force rather than dissipative absorption. Occupation of antibonding Ni $e_g$ orbitals deforms the $\mathrm{NiS}_6$ octahedra, and the 8.0~THz mode, which modulates the Ni--S bonds, tracks this persistent local distortion.
\end{itemize}

\subsubsection{Parity Considerations for Cooperative Phonon-assisted Self-trapping}\label{si:srSTE}
\vspace{2em}

The temperature-dependent localization dynamics yield an activation energy of $E_b \approx 45 \pm 2$ meV, establishing the characteristic energy scale associated with the formation of the localized state. This value exceeds the quantum energy of either the intrinsic 8.0 THz mode ($\hbar\omega_1 \approx 33$ meV) or the transiently activated 2.7 THz mode ($\hbar\omega_2 \approx 11$ meV) individually, but closely matches their combined energy ($\hbar\omega_1+\hbar\omega_2 \approx 44$ meV). This observation motivates consideration of a cooperative multi-phonon pathway for exciton localization. Self-trapping requires the formation of a static, totally symmetric lattice distortion that lowers the electronic energy at the trapping site. This is the symmetry requirement, rather than any particular energy scale, that determines which lattice coordinates can participate. Microscopically, the exciton-lattice interaction is described by the electron-phonon ($e-ph$) Hamiltonian. Expanding the interaction potential with respect to the normal mode lattice coordinates $Q_q$ yields both linear (first-order) and quadratic (second-order) terms\cite{MNBYBook}:

\vspace{0.5em}
\begin{equation}
H_{e-ph} = \sum_{k,q} M_q c_{k+q}^\dagger c_k Q_q + \sum_{k,q_1,q_2} W_{q_1,q_2} c_{k+q_1+q_2}^\dagger c_k Q_{q_1} Q_{q_2}
\end{equation}
\vspace{0.5em}

\noindent where $c^\dagger$ ($c$) creates (annihilates) a charge-transfer excitation, and $M_q$ and $W_{q_1,q_2}$ denote the linear and quadratic electron--phonon coupling coefficients, respectively. Within the $P2_1/m$ subgroup of NiPS$_3$, both the intrinsic 8.0 THz Ni--S stretching mode and the transiently activated 2.7 THz mode transform as odd-parity $A_u$. Because the photoinduced distortion is local and perturbative, this classification remains a valid reference for identifying the modes. For a parity-definite free exciton, the diagonal matrix element associated with an isolated odd-parity distortion vanishes, $\langle \psi | Q_{q}  | \psi \rangle = 0$. Consequently, a single $A_u$ phonon cannot by itself generate the totally symmetric lattice deformation associated with the formation of a static localization potential.

The prominence of the infrared-active $A_u$ phonons nevertheless reflects the nature of the charge-transfer excitation itself. The ligand-to-metal charge-transfer process ($\mathrm{S}\,3p \rightarrow \mathrm{Ni}\,3d$) generates a transient redistribution of charge density and an associated local dipolar character. Odd-parity lattice vibrations efficiently modulate this transient polarization and are therefore expected to couple strongly to the photoexcited state. By contrast, Raman-active $A_g$ phonons correspond to fully symmetric distortions that primarily couple through deformation-potential interactions and do not directly modulate the transient charge-transfer dipole. Within this framework, self-trapping need not be governed by a single phonon coordinate. Instead, theoretical descriptions of self-trapped excitons frequently invoke cooperative lattice distortions involving multiple vibrational modes \cite{STEbook}. Group-theoretical considerations provide a natural illustration of such a mechanism. The direct product of two odd-parity $A_u$ representations transforms as a totally symmetric representation, $A_u \otimes A_u = A_g$. Accordingly, the quadratic coupling channel generates a non-vanishing diagonal matrix element, $\langle \psi | Q_{q_1} Q_{q_2} | \psi \rangle \neq 0$, allowing two odd-parity distortions to cooperate in constructing a symmetry-allowed localization potential. Both modes deform the Ni--S bonds of the $\mathrm{NiS}_6$ octahedra, so the quadratic term acts on the local unit that hosts the charge-transfer exciton. Combining the zone-boundary 2.7 THz mode with the zone-center 8.0 THz mode, this term conserves crystal momentum once localization removes the C-centering translation. 

Parity alone, however, is insufficient to uniquely identify the relevant quadratic channel because the same symmetry condition can also be satisfied by overtone processes $(q_1=q_2)$. The experimentally determined activation barrier ($E_b$) provides the additional constraint. Among the two-phonon combinations of the observed modes, 2.7+8.0~THz, 4.5+5.7~THz, and 2$\times$5.7~THz are energetically compatible with $E_b \approx 45 \pm 2$~meV (Supplementary~\ref{si:1dt}). The remaining combinations lie outside this range. This energy screen defines the competing channels against which the dominant localization pathway can be assessed. The overtone of the 5.7~THz $B_u$ mode ($2\hbar\omega_{B_u} \approx 47$~meV) formally satisfies both the totally symmetric parity requirement ($B_u \otimes B_u = A_g$) and the energetic budget. However, as established by the mode-resolved dynamics in Supplementary~\ref{si:spectro}, the 5.7~THz mode makes a transient, non-dominant contribution to the photoinduced lattice response. By contrast, the 8.0~THz mode exhibits persistent renormalization of the local Ni--S restoring force, while the 2.7~THz response evolves over the localization timescale. The 5.7~THz mode therefore contributes to early energy redistribution, but combinations containing it are unlikely to provide the sustained structural component that dominates localization. For the 4.5+5.7~THz channel, this limitation is reinforced by the weaker activation of the 4.5~THz branch (Supplementary~\ref{si:phonon}). Together with their combined energy, these observations support assignment of the 2.7+8.0~THz pair as the dominant cooperative contribution to the localization coordinate. Within the $C_{2h}$ point group associated with the $P2_1/m$ subgroup, $A_u\otimes A_u=A_g$ provides symmetry support for this cooperative pair.

The weak participation of the 5.7~THz overtone channel may further reflect non-adiabatic coordinate redirection, as suggested by its mode-resolved dielectric signature (Supplementary~\ref{si:spectro}). During the initial wavepacket collapse, cross-coupling between simultaneously driven $A_u$ and $B_u$ coordinates yields an asymmetric shear distortion ($A_u \otimes B_u = B_g$) that provides no polaronic stabilization. The coherent wavepacket along the $B_u$ coordinate may therefore be reflected, consistent with the sub-100~fs kinetic bounce in its $\Delta\varepsilon_{\mathrm{im}}$ response, which decouples the 5.7~THz mode from the subsequent structural relaxation.

%%%%%%%%%%%%%%%% SUPPLEMENTARY FIGURES %%%%%%%%%%%%%%%

%%%%%%%%%%%%%%%% SUPPLEMENTARY REFERENCES %%%%%%%%%%%%%%%

% Do NOT include a reference list in the supplement.
% All references must be in a single list at the end of the main text.

\end{document}